\documentclass{article}

\usepackage{arxiv}

\usepackage[utf8]{inputenc}
\usepackage[T1]{fontenc}
\usepackage[scaled=.98]{XCharter}
\usepackage[type1]{sourcesanspro}
\usepackage[scaled=1.1]{zlmtt}
\usepackage{amsmath}
\usepackage{amssymb}
\usepackage[uprightscript,charter,vvarbb,scaled=1.05]{newtxmath}
\usepackage[hyphens]{url}
\usepackage{xcolor}
\definecolor{LinkBlue}{HTML}{2457C5}
\definecolor{AbstractGray}{HTML}{F3F5F7}
\definecolor{AbstractBorder}{HTML}{DDE3EA}
\usepackage[
  colorlinks=true,
  linkcolor=LinkBlue,
  citecolor=LinkBlue,
  urlcolor=LinkBlue
]{hyperref}
\usepackage{graphicx}
\usepackage{fontawesome5}
\usepackage{booktabs}
\usepackage{microtype}
\usepackage{enumitem}
\usepackage{etoolbox}
\usepackage[numbers,sort&compress]{natbib}
\usepackage{doi}

\usepackage{listings}
\usepackage[most]{tcolorbox}
\tcbuselibrary{breakable,skins}
\lstdefinestyle{promptstyle}{
  basicstyle=\footnotesize\ttfamily,
  breaklines=true,
  columns=fullflexible,
  keepspaces=true,
  showstringspaces=false,
  aboveskip=0pt,
  belowskip=0pt
}
\newtcolorbox{promptbox}[1]{
  breakable,
  enhanced,
  arc=2pt,
  colback=black!2,
  colframe=black!55,
  boxrule=0.5pt,
  left=6pt,
  right=4pt,
  top=2pt,
  bottom=2pt,
  title={#1},
  fonttitle=\bfseries\footnotesize\sffamily,
  colbacktitle=black!12,
  coltitle=black
}

\newtcolorbox{preprintabstract}{
  enhanced,
  breakable,
  colback=AbstractGray,
  colframe=AbstractBorder,
  boxrule=0.35pt,
  arc=3pt,
  left=12pt,
  right=12pt,
  top=5pt,
  bottom=9pt,
  before skip=10pt,
  after skip=16pt,
  title={Abstract},
  fonttitle=\large\bfseries\sffamily,
  coltitle=black,
  colbacktitle=AbstractGray,
  halign title=center,
  titlerule=0pt
}

\renewenvironment{abstract}
  {\begin{preprintabstract}\fontsize{10pt}{12.4pt}\selectfont}
  {\end{preprintabstract}}

\newlist{contributionlist}{itemize}{1}
\setlist[contributionlist]{
  label=\textbullet,
  leftmargin=1.55em,
  labelsep=0.55em,
  itemsep=0.55em,
  topsep=0.55em,
  parsep=0pt,
  partopsep=0pt
}

\pdftrailerid{}

\IfFileExists{evidence-macros.tex}
  {
\newcommand{\NAtcoderAcbBuggy}{589}

\newcommand{\NAtcoderAbcFp}{123}
\newcommand{\NAtcoderAcbNearclone}{578}
\newcommand{\NAtcoderAcbFamilies}{224}
\newcommand{\NAtcoderAcbTopShare}{15.2}
\newcommand{\NAtcoderOracleKThreeRecov}{1.000}
\newcommand{\NAtcoderOracleKOneRecov}{0.925}
\newcommand{\NAtcoderAcbUnionCleanFloor}{906}

\newcommand{\NAtcoderAcbUnionTolFp}{24}
\newcommand{\NAtcoderAcbUnionGateRejected}{1094}
\newcommand{\NAtcoderAcbUnionCleanConfirmed}{905}
\newcommand{\NAtcoderAcbArmCodexLegal}{589}
\newcommand{\NAtcoderAcbArmClaudeLegal}{597}
\newcommand{\NAtcoderAcbArmAgyLegal}{576}
\newcommand{\NAtcoderAcbArmOpencodeLegal}{514}
\newcommand{\NAtcoderAcbArmMiniLegal}{528}
\newcommand{\NAtcoderAcbUnionNew}{317}
\newcommand{\NAtcoderAcbUnionCodexExclusive}{130}
\newcommand{\NAtcoderAcbUnionFdrSampleN}{203}
\newcommand{\NAtcoderAcbUnionFdrCoverageN}{202}
\newcommand{\NAtcoderAcbUnionFdrFp}{0}
\newcommand{\NAtcoderAcbUnionFdrUpperPct}{1.87}
\newcommand{\NCfAdgCovTwenty}{0.853}
\newcommand{\NCfNBuggy}{63}
\newcommand{\NTceAgentCovTwenty}{0.838}
\newcommand{\NTceSatOneRandom}{0.65}
\newcommand{\NTceSatOneAgent}{0.32}
\newcommand{\NTceMargRandom}{0.004}
\newcommand{\NTceMargAgent}{0.026}
\newcommand{\NTceMargAgentNofb}{0.025}
\newcommand{\NTceMargRatio}{5.8}
\newcommand{\NAtcoderRejCovCodex}{0.930}
\newcommand{\NAtcoderRejCovClaude}{0.932}
\newcommand{\NAtcoderRejCovAgy}{0.925}
\newcommand{\NAtcoderRejCovOpencode}{0.922}
\newcommand{\NAtcoderRejCovMini}{0.919}
\newcommand{\NAtcoderRejOfficialCov}{0.936}
\newcommand{\NAtcoderRejRandomCov}{0.827}
\newcommand{\NAtcoderRejAgentCovBest}{0.932}
\newcommand{\NAtcoderRejAgentCovWorst}{0.919}
\newcommand{\NAtcoderRejAgentWorstGap}{1.7}
\newcommand{\NAtcoderRejRandomGap}{10.9}
\newcommand{\NAtcoderRejSelfgapRows}{1696}
\newcommand{\NAtcoderRejSelfgapPct}{6.4}
\newcommand{\NAtcoderRejSelfgapTleShare}{90.9}
\newcommand{\NAtcoderRejSelfgapDrift}{154}
\newcommand{\NAtcoderRejSelfgapDriftPct}{0.58}
\newcommand{\NAtcoderRejParityGapReproPp}{2.1}

\newcommand{\NAtcoderAcUnionCov}{0.964}
\newcommand{\NAtcoderAcOnlyOfficialUnion}{41}
\newcommand{\NAtcoderAcPanelPool}{4013}

\newcommand{\NAtcoderNiCiNoninferior}{2}
\newcommand{\NAtcoderNiHolmPass}{0}
\newcommand{\NAtcoderNiHolmPassConfirmatory}{2}

\newcommand{\NCfCovOfficial}{0.000}
\newcommand{\NCfCovRandom}{0.377}
\newcommand{\NCfCovCodecontests}{0.403}
\newcommand{\NCfCovCcplus}{0.725}
\newcommand{\NCfCovEvalplus}{0.625}
\newcommand{\NCfAdgCcplusDelta}{0.129}
\newcommand{\NCfAdgCcplusCiLo}{-0.011}
\newcommand{\NCfAdgCcplusCiHi}{0.351}
\newcommand{\NCfAdgCcplusDeltaFifty}{0.143}
\newcommand{\NCfAdgCcplusCiLoFifty}{0.028}
\newcommand{\NCfAdgCcplusCiHiFifty}{0.331}

\newcommand{\NAtcoderOracleKFiveRecov}{1.000}

\newcommand{\NAtcoderNiDeltaLoCodex}{-0.017}
\newcommand{\NAtcoderNiDeltaLoClaude}{-0.018}
\newcommand{\NAtcoderNiDeltaLoAgy}{-0.027}
\newcommand{\NAtcoderNiDeltaLoOpencode}{-0.035}
\newcommand{\NAtcoderNiDeltaLoMini}{-0.036}

\newcommand{\NTceMargRandomSe}{0.0013}
\newcommand{\NTceMargAgentSe}{0.0046}
\newcommand{\NTceMargAgentNofbSe}{0.0043}
\newcommand{\NAtcoderRefCount}{848}
\newcommand{\NAtcoderRefNearcloneClusters}{848}
\newcommand{\NAtcoderNiDeltaCodex}{-0.004}
\newcommand{\NAtcoderNiDeltaHiCodex}{0.009}
\newcommand{\NAtcoderNiDeltaClaude}{-0.003}
\newcommand{\NAtcoderNiDeltaHiClaude}{0.011}
\newcommand{\NAtcoderNiDeltaAgy}{-0.011}
\newcommand{\NAtcoderNiDeltaHiAgy}{0.005}
\newcommand{\NAtcoderNiDeltaOpencode}{-0.013}
\newcommand{\NAtcoderNiDeltaHiOpencode}{0.007}
\newcommand{\NAtcoderNiDeltaMini}{-0.015}
\newcommand{\NAtcoderNiDeltaHiMini}{0.003}

\newcommand{\NAtcoderRejUnionPool}{26,682}
\newcommand{\NAtcoderAuditedAcTotal}{20,375}
\newcommand{\NAtcoderAcbProblems}{74}
\newcommand{\NAtcoderTolFp}{15}
\newcommand{\NAtcoderConsensusFp}{125}
\newcommand{\NAtcoderConsensusVoterSample}{80}
\newcommand{\NAtcoderAcbOfficialTleCaught}{10}
\newcommand{\NAtcoderAcbLedgerCandidates}{736}
\newcommand{\NAtcoderAcbNondeterministic}{7}
\newcommand{\NAtcoderAcbRawDet}{729}
\newcommand{\NAtcoderAuditedSharePct}{8.9}
\newcommand{\NCfOracleGoldUsable}{1,795}
\newcommand{\NCfOracleContestedTotal}{15}
\newcommand{\NCfOracleContestedTop}{13}
\newcommand{\NCfOracleContestedOther}{2}
\newcommand{\NAtcoderKillInputLegal}{589}
\newcommand{\NCfKillmatrixValApplied}{41}
\newcommand{\NCfProblems}{41}
\newcommand{\NCfAdgLeadMinGap}{0.018}
\newcommand{\NAtcoderAcbRate}{2.9}
\newcommand{\NAtcoderProblemsTotal}{106}
\newcommand{\NAtcoderAcbWa}{545}
\newcommand{\NAtcoderAcbRe}{44}
\newcommand{\NAtcoderOnlyOfficialExcl}{1136}
\newcommand{\NAtcoderOnlyAdgExcl}{746}
\newcommand{\NAtcoderOracleRefsPerProblem}{8}

\newcommand{\NAtcoderESevenOverlapMaxPct}{5.5}
\newcommand{\NCfAdgInputsMean}{43.8}
\newcommand{\NCfBaselineInputsMin}{49.0}
\newcommand{\NCfBaselineInputsMax}{54.1}
\newcommand{\NCfAdgFinalPerProblem}{50}
\newcommand{\NTceProblems}{24}
\newcommand{\NTceBandLo}{1500}
\newcommand{\NTceBandHiLo}{3500}
\newcommand{\NCfCovOneOfficial}{0.000}
\newcommand{\NCfCovOneRandom}{0.175}
\newcommand{\NCfCovOneCodecontests}{0.075}
\newcommand{\NCfCovOneCcplus}{0.205}
\newcommand{\NCfCovOneEvalplus}{0.123}
\newcommand{\NCfAdgCovOne}{0.223}
\newcommand{\NCfCovTwoOfficial}{0.000}
\newcommand{\NCfCovTwoRandom}{0.211}
\newcommand{\NCfCovTwoCodecontests}{0.134}
\newcommand{\NCfCovTwoCcplus}{0.336}
\newcommand{\NCfCovTwoEvalplus}{0.209}
\newcommand{\NCfAdgCovTwo}{0.356}
\newcommand{\NCfCovFiveOfficial}{0.000}
\newcommand{\NCfCovFiveRandom}{0.265}
\newcommand{\NCfCovFiveCodecontests}{0.246}
\newcommand{\NCfCovFiveCcplus}{0.531}
\newcommand{\NCfCovFiveEvalplus}{0.377}
\newcommand{\NCfAdgCovFive}{0.570}
\newcommand{\NCfCovTenOfficial}{0.000}
\newcommand{\NCfCovTenRandom}{0.316}
\newcommand{\NCfCovTenCodecontests}{0.332}
\newcommand{\NCfCovTenCcplus}{0.644}
\newcommand{\NCfCovTenEvalplus}{0.520}
\newcommand{\NCfAdgCovTen}{0.729}
\newcommand{\NCfCovFortyOfficial}{0.000}
\newcommand{\NCfCovFortyRandom}{0.437}
\newcommand{\NCfCovFortyCodecontests}{0.481}
\newcommand{\NCfCovFortyCcplus}{0.790}
\newcommand{\NCfCovFortyEvalplus}{0.691}
\newcommand{\NCfAdgCovForty}{0.932}
\newcommand{\NCfCovSixtyOfficial}{0.000}
\newcommand{\NCfCovSixtyRandom}{0.460}
\newcommand{\NCfCovSixtyCodecontests}{0.524}
\newcommand{\NCfCovSixtyCcplus}{0.825}
\newcommand{\NCfCovSixtyEvalplus}{0.730}
\newcommand{\NCfAdgCovSixty}{0.952}
\newcommand{\NCfCovFiftyOfficial}{0.000}
\newcommand{\NCfCovFiftyRandom}{0.453}
\newcommand{\NCfCovFiftyCodecontests}{0.505}
\newcommand{\NCfCovFiftyCcplus}{0.809}
\newcommand{\NCfCovFiftyEvalplus}{0.711}
\newcommand{\NCfAdgCovFifty}{0.952}

\newcommand{\NCfInputsOfficial}{1.0}
\newcommand{\NCfInputsRandom}{49.0}
\newcommand{\NCfInputsCodecontests}{54.1}
\newcommand{\NCfInputsCcplus}{53.2}
\newcommand{\NCfInputsEvalplus}{54.1}

\newcommand{\NAtcoderBugclassSpecialCaseMissing}{97}
\newcommand{\NAtcoderBugclassWrongGreedy}{73}
\newcommand{\NAtcoderBugclassOffByOne}{64}
\newcommand{\NAtcoderBugclassIntegerOverflow}{62}
\newcommand{\NAtcoderBugclassArrayOutOfBounds}{55}
\newcommand{\NAtcoderBugclassBoundary}{55}
\newcommand{\NAtcoderBugclassTieBreaking}{33}
\newcommand{\NAtcoderBugclassWrongAlgorithm}{27}
\newcommand{\NAtcoderBugclassPrecision}{23}
\newcommand{\NAtcoderBugclassOther}{100}
\newcommand{\NAtcoderBugclassOtherClasses}{37}
\newcommand{\NAtcoderBugclassDistinct}{46}
\newcommand{\NAtcoderCostOutKtokCodex}{30.5}
\newcommand{\NAtcoderCostOutKtokClaude}{21.7}
\newcommand{\NAtcoderCostOutKtokOpencode}{26.3}
\newcommand{\NAtcoderCostWallMinCodex}{12.9}
\newcommand{\NAtcoderCostWallMinClaude}{5.8}
\newcommand{\NAtcoderCostWallMinOpencode}{22.8}
\newcommand{\NAtcoderCostWallMinMini}{9.1}
\newcommand{\NCfCostOutKtok}{27.3}
\newcommand{\NCfCostWallMin}{7.7}
\newcommand{\NCfCostTurns}{30}
\newcommand{\NAtcoderBugclassAgreementN}{60}
\newcommand{\NAtcoderBugclassAgreementPct}{43.3}
\newcommand{\NAtcoderBugclassAgreementKappa}{0.36}
\newcommand{\NAtcoderTrajSemanticAdv}{106}
\newcommand{\NAtcoderTrajStructural}{104}
\newcommand{\NAtcoderTrajSweep}{52}
\newcommand{\NAtcoderTrajEnum}{18}
\newcommand{\NAtcoderTrajSelfWrongPanel}{106}
\newcommand{\NAtcoderTrajPairedMining}{103}
\newcommand{\NAtcoderTrajMultiTwo}{105}
\newcommand{\NAtcoderTrajAgreeSemanticAdvPct}{100.0}
\newcommand{\NAtcoderTrajAgreeStructuralPct}{97.2}
\newcommand{\NAtcoderTrajAgreeSweepPct}{84.9}
\newcommand{\NAtcoderTrajAgreeEnumPct}{50.0}
\newcommand{\NAtcoderTrajAgreeSelfWrongPanelPct}{100.0}
\newcommand{\NAtcoderTrajAgreePairedMiningPct}{97.2}
\newcommand{\NAtcoderTrajKappaSemanticAdv}{1.00}
\newcommand{\NAtcoderTrajKappaStructural}{-0.01}
\newcommand{\NAtcoderTrajKappaSweep}{0.70}
\newcommand{\NAtcoderTrajKappaEnum}{0.15}
\newcommand{\NAtcoderTrajKappaSelfWrongPanel}{1.00}
\newcommand{\NAtcoderTrajKappaPairedMining}{0.00}

\newcommand{\NAtcoderOracleCensusN}{589}
\newcommand{\NAtcoderOracleCensusFp}{0}
\newcommand{\NAtcoderCensusSolverInputs}{188}
\newcommand{\NAtcoderOracleSampleFdrUpperPct}{2.3}
\newcommand{\NAtcoderSelectionCandidates}{126}
\newcommand{\NAtcoderSelectionEligible}{106}
\newcommand{\NCfSolverPoolSingle}{1,640}
\newcommand{\NCfSolverPoolAgent}{324}
\newcommand{\NCfSolverPoolTotal}{1,964}
\newcommand{\NCfSamplePassing}{596}
\newcommand{\NCfHarvestProblems}{48}
\newcommand{\NCfExcludedProblems}{7}
\newcommand{\NAtcoderValidatorFuzzFalseAccepts}{0}
\newcommand{\NAtcoderValidatorFuzzInputs}{354}
\newcommand{\NAtcoderValidatorFuzzProblems}{25}
\newcommand{\NAtcoderValidatorSemanticFalseAccepts}{0}
\newcommand{\NAtcoderValidatorSemanticInputs}{62}
\newcommand{\NAtcoderValidatorSemanticProblems}{18}
\newcommand{\NAtcoderLeakAuditAccessLeaks}{0}

\newcommand{\NAtcoderLeakAuditScannedRuns}{529}
\newcommand{\NCfTopupTarget}{60}
\newcommand{\NCfTopupProblems}{27}
\newcommand{\NCfTopupNativeCc}{14}
\newcommand{\NCfTopupNativeEp}{39}
\newcommand{\NCfTopupNativeRandom}{50}
\newcommand{\NAtcoderRejLegalityProblems}{103}
\newcommand{\NAtcoderRejLegalityMinPct}{0.08}
\newcommand{\NAtcoderRejLegalityMaxPct}{3.79}
\newcommand{\NAtcoderRejLegalityRandomPct}{1.67}
\newcommand{\NCfBestofnNBuggy}{62}
\newcommand{\NCfBestofnNProblems}{19}
\newcommand{\NCfBestofnAgentCovFifty}{0.903}
\newcommand{\NCfBestofnAgentMissFifty}{0.097}
\newcommand{\NCfBestofnAgentCovTwenty}{0.801}

\newcommand{\NCfBestofnSingleCovFifty}{0.665}
\newcommand{\NCfBestofnSingleMissFifty}{0.335}
\newcommand{\NCfBestofnSingleMissRatioFifty}{3.46}
\newcommand{\NCfBestofnSingleDeltaFifty}{0.238}
\newcommand{\NCfBestofnSingleCiLoFifty}{0.074}
\newcommand{\NCfBestofnSingleCiHiFifty}{0.417}
\newcommand{\NCfBestofnSingleCovTwenty}{0.597}

\newcommand{\NCfBestofnSingleDeltaTwenty}{0.204}
\newcommand{\NCfBestofnSingleCiLoTwenty}{0.059}
\newcommand{\NCfBestofnSingleCiHiTwenty}{0.343}
\newcommand{\NCfBestofnUnionCovFifty}{0.790}
\newcommand{\NCfBestofnUnionMissFifty}{0.210}

\newcommand{\NCfBestofnUnionDeltaFifty}{0.113}
\newcommand{\NCfBestofnUnionCiLoFifty}{-0.102}
\newcommand{\NCfBestofnUnionCiHiFifty}{0.339}
\newcommand{\NCfBestofnUnionCovTwenty}{0.688}

\newcommand{\NCfBestofnUnionDeltaTwenty}{0.113}
\newcommand{\NCfBestofnUnionCiLoTwenty}{-0.070}
\newcommand{\NCfBestofnUnionCiHiTwenty}{0.298}
\newcommand{\NCfBestofnOracleTwoCovFifty}{0.785}
\newcommand{\NCfBestofnOracleThreeCovFifty}{0.831}

\newcommand{\NCfBestofnOracleFourCovFifty}{0.855}

\newcommand{\NCfBestofnOracleFourDeltaFifty}{0.048}
\newcommand{\NCfBestofnOracleFourCiLoFifty}{-0.130}
\newcommand{\NCfBestofnOracleFourCiHiFifty}{0.212}

\newcommand{\NCfBestofnAgentCovOne}{0.168}
\newcommand{\NCfBestofnOracleFourCovOne}{0.346}
\newcommand{\NCfBestofnOracleFourDeltaOne}{-0.178}
\newcommand{\NCfBestofnOracleFourCiLoOne}{-0.339}
\newcommand{\NCfBestofnOracleFourCiHiOne}{-0.031}
\newcommand{\NCfBestofnMixtureGapClosedFifty}{53}
\newcommand{\NCfBestofnAgentLegalRate}{0.995}
\newcommand{\NCfBestofnSingleLegalRateMin}{0.917}
\newcommand{\NCfBestofnSingleLegalRateMax}{0.981}
\newcommand{\NCfBestofnSingleZeroLegalRepProblems}{3}

\newcommand{\NCfMissRatioCcplusFifty}{4.01}
\newcommand{\NCfTypemixAgentMaxsizeRate}{0.18}
\newcommand{\NCfTypemixSingleMaxsizeRateMin}{0.41}
\newcommand{\NCfTypemixSingleMaxsizeRateMax}{0.55}

\newcommand{\NCfBestofnCcplusDsMeanMissFifty}{0.586}
\newcommand{\NCfBestofnCcplusDsMeanCovFifty}{0.414}
\newcommand{\NCfBestofnCcplusDsMeanCovTwenty}{0.362}
\newcommand{\NCfBestofnCcplusDsMissRatioFifty}{6.05}
\newcommand{\NCfBestofnCcplusDsDeltaFifty}{+0.489}
\newcommand{\NCfBestofnCcplusDsCiLoFifty}{+0.359}
\newcommand{\NCfBestofnCcplusDsCiHiFifty}{+0.623}
\newcommand{\NCfBestofnCcplusDsDeltaTwenty}{+0.439}
\newcommand{\NCfBestofnCcplusDsCiLoTwenty}{+0.310}
\newcommand{\NCfBestofnCcplusDsCiHiTwenty}{+0.552}
\newcommand{\NCfBestofnCcplusDsRepOneCovFifty}{0.158}
\newcommand{\NCfBestofnCcplusDsRepTwoCovFifty}{0.562}
\newcommand{\NCfBestofnCcplusDsRepThreeCovFifty}{0.677}
\newcommand{\NCfBestofnCcplusDsRepFourCovFifty}{0.259}
\newcommand{\NCfBestofnCcplusDsGvInconsistent}{4}
\newcommand{\NCfBestofnCcplusGvInconsistentOpus}{0}
}
  {\IfFileExists{../../evidence/derived/evidence-macros.tex}
    {}{}}

\providecommand{\evclaim}[1]{}
\providecommand{\evfact}[1]{}

\IfFileExists{sections/01-introduction.tex}
  {}{}

\newcommand{\mainpapername}{main text}
\newcommand{\mainpapersection}[1]{Section~#1}

\newcommand{\overviewfigurewidth}{0.94\linewidth}
\newcommand{\mainplotwidth}{0.62\linewidth}
\newcommand{\appendixplotwidth}{0.62\linewidth}
\newcommand{\fitstrategytable}[1]{#1}
\newcommand{\appendixwidetablepadding}{\setlength{\tabcolsep}{4.5pt}}

\graphicspath{{figures/}{../../figures/}}

\newcommand{\papertitle}{Coding Agents as Test-Suite Auditors:\\
Finding What Official Suites Miss While Approaching What They Catch}
\newcommand{\papershorttitle}{Coding Agents as Test-Suite Auditors}
\newcommand{\paperstatus}{Preprint}

\newcommand{\correspondenceemail}{cswmzuo@gmail.com}
\newcommand{\artifacturl}{https://github.com/xieTwim/test-suite-auditors}
\newcommand{\artifactdisplay}{github.com/xieTwim/test-suite-auditors}
\newcommand{\companionurl}{https://github.com/xieTwim/agentic-testgen}
\newcommand{\companiondisplay}{github.com/xieTwim/agentic-testgen}

\title{\normalfont\sffamily\bfseries\papertitle}

\newcommand{\authornamefont}{\fontsize{11.8pt}{14.2pt}\selectfont\sffamily\bfseries}
\newcommand{\affiliationfont}{\fontsize{10.2pt}{12.4pt}\selectfont\normalfont}
\newlength{\authorrowbreakheight}
\makeatletter
\patchcmd{\@maketitle}
  {\begin{tabular}[t]{c}\bf\rule{\z@}{24\p@}\ignorespaces}
  {\begin{tabular}[t]{c}\authornamefont\rule{\z@}{24\p@}\ignorespaces}
 {}{\PackageWarning{adg-arxiv}{Could not patch the first author-row break}}
\patchcmd{\@maketitle}
  {\begin{tabular}[t]{c}\bf\rule{\z@}{24\p@}\ignorespaces}
  {\begin{tabular}[t]{c}\authornamefont\rule{\z@}{\authorrowbreakheight}\ignorespaces}
  {}{\PackageWarning{adg-arxiv}{Could not patch the second author-row break}}
\makeatother
\newcommand{\resourcelink}[2]{%
  \href{#1}{\normalfont\mdseries\nolinkurl{#2}}%
}
\newcommand{\resourceentry}[4]{%
  \raisebox{-0.05ex}{\makebox[1.35em][c]{\normalsize #1}}&
  \textbf{#2:}\ \resourcelink{#3}{#4}%
}

\author{
  \authornamefont
  \textbf{Shuyang Xie}\textsuperscript{1}\thanks{Equal contribution.}
  \quad
  \textbf{Shuxiao Xie}\textsuperscript{2}\footnotemark[1]
  \quad
  \textbf{Feng Zhu}\textsuperscript{1}
  \quad
  \textbf{Yanli Ji}\textsuperscript{3}
  \quad
  \textbf{Wangmeng Zuo}\textsuperscript{1}\thanks{Corresponding author:
  \resourcelink{mailto:\correspondenceemail}{\correspondenceemail}.}
  \\[0.7em]
  \affiliationfont
  \textsuperscript{1}Harbin Institute of Technology
  \quad
  \textsuperscript{2}Fudan University
  \quad
  \textsuperscript{3}Sun Yat-sen University
}

\date{\small
\begin{tabular}{@{}r@{\hspace{0.4em}}l@{}}
\resourceentry{\faGithub}{Research artifacts}{\artifacturl}{\artifactdisplay}
\\[0.15em]
\resourceentry{\faGithub}{Companion skill}{\companionurl}{\companiondisplay}
\end{tabular}
}

\renewcommand{\headeright}{\paperstatus}
\renewcommand{\undertitle}{\paperstatus}
\renewcommand{\shorttitle}{\papershorttitle}

\hypersetup{
  pdftitle={Coding Agents as Test-Suite Auditors: Finding What Official Suites Miss While Approaching What They Catch},
  pdfsubject={Public preprint},
  pdfauthor={Shuyang Xie, Shuxiao Xie, Feng Zhu, Yanli Ji, Wangmeng Zuo}
}

\begin{document}
\maketitle

\begin{abstract}
\looseness=-1
Online-judge verdicts and the datasets and benchmarks built on them are treated as ground
truth for evaluating and training large language models for code. Yet prior audits have
sounded a warning: official suites accept buggy submissions. These audits, however, stop at
the warning and offer no practical remedy. Our remedy has two parts: an off-the-shelf coding
agent, serving as a test-suite auditor, both builds adversarial test suites to expose what official
suites miss and supplies these suites where no official suite exists; a certification chain determines
whether each agent-flagged submission is genuinely buggy without relying on the official
judge: multiple independently written accepted solutions agree on the expected output for
every test, brute-force solutions settle disagreements, and a per-problem validator certifies
each failing input legal.
One such agent identifies \NAtcoderAcbBuggy{} verified accepted-but-buggy submissions among AtCoder's
\NAtcoderAuditedAcTotal{} audited accepted submissions; extending the same certification to all five
agents yields a union floor of \NAtcoderAcbUnionCleanFloor{} such submissions. Five agents, scored
separately, each stay within \NAtcoderRejAgentWorstGap{}pp of official-suite coverage on logic bugs those
suites catch.
On post-cutoff Codeforces problems with no available official suites, the same test-building method
leads all five reproduced baselines at every tested input budget. Where an official suite exists, the agent audits suite adequacy
instead of assuming it; where none exists, agent suites catch the most buggy submissions
among methods we reproduced and tested.
\end{abstract}

\section{Introduction}
\label{sec:intro}

\looseness=-1
Online judges make hidden test suites the operational ground truth for program correctness. A passing
submission is marked \emph{accepted}, so the verdict shapes contestant ratings, determines which
solutions datasets and benchmarks treat as correct, and grounds the execution-based reward signals
used to train code models \citep{li-competitionlevel-2022,li-taco-2023}. Yet the trust chain can
fail: a solution can pass every official test yet fail on a legal input the suite never tries.

\looseness=-1
That risk is no longer hypothetical. An empirical audit found official AtCoder tests accepting buggy
submissions at scale \citep{liu-whojudges-2023}, and TrickyBugs assembled plausible-but-buggy
programs into a dataset \citep{liu-trickybugs-2024}. Together, these studies establish the blind
spot through random and differential testing over graded submissions, with majority voting used for
adjudication. Downstream datasets inherit the same blind spot when they treat passage of their own
tests as correctness. Without official suites, CodeContests and TACO ship tests produced by mutation
or large language model generation. In those shipped tests, illegal, out-of-spec inputs are a named
failure mode \citep{wang-codecontests-2025}. Prior work builds tests for datasets, evaluates test
writers on known bugs, or audits judges through graded submissions (\S\ref{sec:background}). No
practical remedy yet combines test generation strong enough to expose what official suites miss
with per-finding certification independent of the official judge (\S\ref{sec:background};
\S\ref{sec:why}).

\begin{figure*}[t]
\centering
\providecommand{\overviewfigurewidth}{\textwidth}
\includegraphics[width=\overviewfigurewidth]{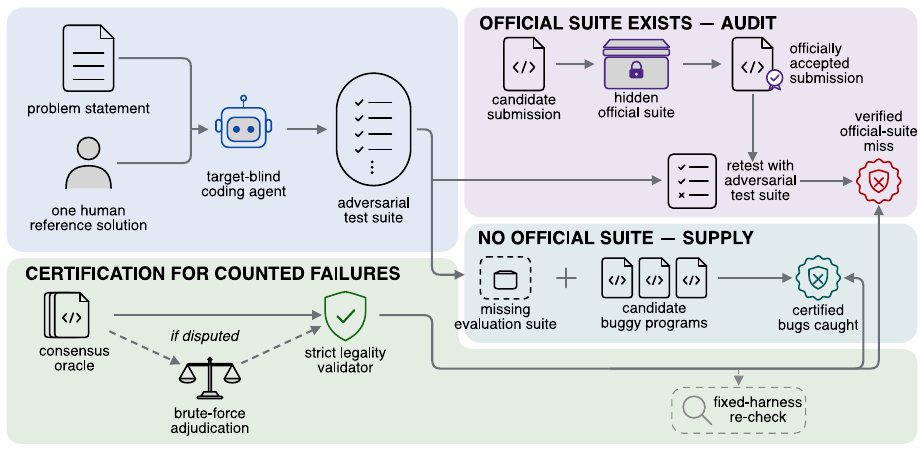}
\caption{The auditor framework and its certification chain (\S\ref{sec:method}). Agents see
only the problem statement and one reference solution; the adversarial test suite audits the
hidden official suite (the audited object, never an input) by retesting an officially accepted
submission, or supplies the missing evaluation suite for candidate buggy programs; the
certification chain serves both.}
\label{fig:overview}
\end{figure*}

\looseness=-1
Off-the-shelf coding agents fill this gap as \emph{test-suite auditors}. Unlike those approaches,
our target-blind method audits the official hidden suites. Each agent receives only a problem
statement and one human reference solution, from which it constructs an adversarial suite; official
tests, verdicts, and the submissions under audit remain hidden. A finding enters the ledger only
after the certification chain in Figure~\ref{fig:overview}: a consensus oracle over independent
accepted solutions supplies the expected output, brute-force solutions settle disputes, and a strict
per-problem input-legality validator written from the statement alone admits as certifying evidence
only legal inputs that pass it. The gate is stated in \S\ref{sec:method} and measured in
\S\ref{sec:why}. Within our own certification chains, it closes the illegal-input failure mode seen
in shipped dataset tests.

\looseness=-1
The same agents generate tests in both roles. Where official suites exist, they audit them: on AtCoder, one agent arm
exposes verified accepted-but-buggy submissions within the audited sample. Every engine
approaches re-judged official-suite logic-bug coverage, complementing the official suites rather than replacing them.
Where none exist, they supply the tests themselves: on fresh post-cutoff Codeforces problems, they
lead every reproduced baseline. The certification chain keeps each finding checkable in both
settings. Our contributions are:

\ifdefined\contributionlist\else
  \newenvironment{contributionlist}{\begin{itemize}}{\end{itemize}}
\fi
\begin{contributionlist}
\looseness=-1\item \textbf{A framework for auditing suite adequacy.} Our \emph{target-blind} auditor turns suite
adequacy from an assumption into an audited property. Prior random and differential testing targets graded
submissions \citep{liu-whojudges-2023,liu-trickybugs-2024}; off-the-shelf coding agents never see the audited
submissions and construct adversarial suites from a problem statement and one reference solution. The certification
chain in \S\ref{sec:method} verifies each finding without relying on the official judge.
\looseness=-1\item \textbf{An AtCoder audit with robustness checks.} We audit
\NAtcoderAuditedAcTotal{} compilable accepted submissions. One agent arm (\textsc{codex}) identifies new, individually verified
accepted-but-buggy official-suite misses; the \NAtcoderAcbBuggy-entry ledger records them.
Carrying the same certification chain across all five arms yields a legality- and tolerance-clean union
floor of \NAtcoderAcbUnionCleanFloor{}.
Separately, on the known-bug set (the official suite's historically rejected submissions), all agent
suites trail the re-judged official detection rate by at most \NAtcoderRejAgentWorstGap{}pp; the
random generation matched to the agent's input budget trails by \NAtcoderRejRandomGap{}pp. Oracle and input audits find no
measurement artifact (\S\ref{sec:why}).
\looseness=-1\item \textbf{Codeforces test supply.} On post-cutoff Codeforces problems
without official suites, agent suites supply the tests. At every tested input budget, the agent arm leads all five baselines
(\S\ref{sec:exp2}). At its design budget of 50 inputs, cov@50 (the buggy-pool share a random 50-input subset kills) is \NCfAdgCovFifty{}.
Among the generator arms, it requires the fewest inputs to reach the coverage it achieves.
\looseness=-1\item \textbf{Open artifacts.} We release the five-arm ledger of accepted-but-buggy
submissions, each entry recording which agents certified it (\NAtcoderAcbUnionCleanFloor{} in union,
\NAtcoderAcbBuggy{} for \textsc{codex} alone), the six-arm CF-fresh kill matrix, the per-problem
validators, and the judging harness. These artifacts support machine re-checking of every certified
number derived from them.
\end{contributionlist}

\section{Background and Related Work}
\label{sec:background}

\looseness=-1
Most prior work treats official suites or platform verdicts as scoring standards, either by building
substitutes or by evaluating whether public and generated tests are sufficient. A smaller line
instead audits the official verdict, but existing approaches rely on uncertified testing over graded
submissions, target-aware inspection, or one-submission-at-a-time attacks. We therefore proceed from
suite construction and test-quality evaluation, to verdict auditing, and finally to input legality
as a certificate.

\paragraph{Building suites and evaluating test quality.}
\looseness=-1
When hidden suites are unavailable, dataset builders construct substitutes. Mutation-based and
inherited suites include CodeContests \citep{li-competitionlevel-2022} and TACO
\citep{li-taco-2023};
agentic and scaled synthesis include CodeContests+, HardTests, and Klear-CodeTest
\citep{wang-codecontests-2025,he-hardtests-2025,fu-klearcodetest-2025}; generated or adversarial
pipelines include AutoCode, Themis, and CodeHacker
\citep{zhou-autocode-2025,ye-themis-2026,shi-codehacker-2026}. A separate line evaluates whether
available tests expose bugs: HumanEval and EvalPlus study public benchmark tests
\citep{chen-evaluating-2021,liu-your-2023}, while TestCase-Eval evaluates LLM tests and TCGBench
their \emph{generators}, and CodeT and CURE use generated tests for solution selection
\citep{yang-can-2025,cao-can-2025,chen-codet-2022,wang-coevolving-2025}. These test-quality studies
assess public or generated tests rather than hidden platform suites. Across both sublines, an
official suite or verdict remains a scoring standard rather than an audit object.

\paragraph{Auditing the official verdict.}
\looseness=-1
Changing the audit object also changes what each system accepts as truth. Who-Judges-the-Judge
directly audits official tests and exposes accepted-but-buggy submissions at scale
\citep{liu-whojudges-2023}. It uses random, differential, and majority-vote testing over graded
submissions, but does not certify its findings. TrickyBugs collects such programs
\citep{liu-trickybugs-2024}, while its LLM-powered follow-up generates tests and evaluates
failures on curated datasets using canonical programs and dataset-provided or manual
input-validity checks \citep{liu-llmpowered-2024}. UOJ-Bench is target-aware: each attack targets one full-score submission, and its cost analysis
prices that per-submission strategy at \$100k/year \citep{xu-uojbench-2026}. Solvita is also
target-aware; it inspects each candidate source and treats disagreement with official acceptance as
an internal diagnostic, with stronger-than-official confirmation relying on accepted-solver
cross-checks and manual validation \citep{li-solvita-2026}. CodeContests-O iterates generation
through execution feedback on per-problem, platform-labeled solution pools and takes the official
verdict as ground truth \citep{cai-codecontestso-2026}; neither those pools nor its shipped tests
cover the post-cutoff problems in \S\ref{sec:exp2}. Under that convention, a generated test killing
a latent accepted-but-buggy solution is a false negative, so the loop repairs the test rather than
treating the disagreement as evidence. We instead audit that verdict.

\paragraph{Making input legality a certificate.}
\looseness=-1
Input legality places a separate requirement on the tests used to audit an official verdict. A
killing input is admissible only when \emph{legal} under the statement constraints. LogiCase makes
validators first-class through formal grammars \citep{sung-logicase-2025}; CodeContests+ identifies
illegal inputs as a failure mode in shipped dataset tests \citep{wang-codecontests-2025}. Themis uses
constraint-aware validators, whereas our per-problem validator is statement-derived and written
while blind to the killing inputs at stake; under our protocol an input serves as evidence only
after passing it. This requirement addresses the illegal-input
failure mode identified in shipped dataset tests. Those tests provide the background for the
requirement; the official verdict remains the audit object. \S\ref{sec:why} measures the statement-derived
gate on the certified kills. Taken together, these distinctions make the audit
protocol, rather than suite scale or platform infrastructure, the basis of our comparison
(\S\ref{sec:method}).

\section{Method: Agents as Suite Auditors, and the Certification Chain}
\label{sec:method}

\looseness=-1
We next ask how a disagreement becomes a certified finding. Every reported finding follows a certification chain with independently audited links (Figure~\ref{fig:overview}). Both experiments (\S\ref{sec:exp1}--\ref{sec:exp2}) instantiate the same certification pattern, with the oracle, engine composition, and validator source configured per experiment.

\paragraph{What the auditor can see.}
\looseness=-1
Suite construction uses a restricted visibility model, not a certification-chain link. For one competitive-programming problem, construction sees only the statement and \emph{one}
human reference solution. It cannot access official tests, verdicts, or any submission under audit. Per-problem artifacts comprise a test generator, an input validator, an adversarial suite of
concrete inputs, and re-checking tooling. Official suites and verdicts enter only afterwards as the \emph{object} audited and are not
used to judge our artifacts. The remaining leakage surface is overlap between constructed inputs and official tests, audited directly in \S\ref{sec:threats}.

\paragraph{How the auditor constructs a suite.}
\looseness=-1
Five off-the-shelf coding-agent engines provide the auditing arms: \textsc{codex}: OpenAI Codex CLI~\citep{openai-codex-2025} on GPT-5.4~\citep{openai-gpt54-2026}; \textsc{claude}: Claude Code~\citep{anthropic-claudecode-2025} on Claude Opus 4.8~\citep{anthropic-opus48-2026}; \textsc{agy}: Antigravity CLI~\citep{google-antigravity-2025} on Gemini 3.1 Pro~\citep{google-gemini31-2026}; \textsc{opencode}: OpenCode CLI~\citep{opencode-2025} on DeepSeek V4 Pro~\citep{deepseek-v4-2026}; and \textsc{mini}: mini-SWE-agent~\citep{yang-sweagent-2024} on DeepSeek V4 Pro under a minimal scaffold. The task brief is identical across engines, and per-engine reasoning-effort settings are in \S A. CodeContests+ fixes two generation--validation pipeline agent roles
\citep{wang-codecontests-2025}. Each engine chooses its construction strategy.

\paragraph{What makes the comparisons fair.}
\looseness=-1
AtCoder evaluates a five-engine panel, but its certified ledger uses one arm (\textsc{codex}); CF-fresh, the post-cutoff setting, runs a single engine. No pooling occurs across compositions. Construction strategy may vary, but every evaluated suite is scored with the same judging harness and scoring basis. Experiment~1 compares agent-constructed and re-judged official suites with random generation
matched to the agent's input budget. Experiment~2 compares agent-constructed suites with five reproduced baselines on a single matrix of output disagreements and scores only validator-passing inputs, pairing scores per problem. The mechanism analysis gives random the same input budget (\S\ref{sec:why}).

\paragraph{How a disagreement becomes a certificate.}
\looseness=-1
A shared scoring basis makes an output disagreement comparable, but not yet a bug. A test input \emph{kills} a submission when the submission's output disagrees with the expected output. The expected output comes from a consensus oracle over independently accepted human solutions. On AtCoder, agreement among \NAtcoderOracleRefsPerProblem{} reference solutions per problem defines the expected output; on Codeforces, agreement among the outputs of all successful runs, after all three accepted reference solutions are run, defines the expected output, provided that at least two runs succeed. \S\ref{sec:why} audits AtCoder-oracle conservatism and Codeforces-oracle soundness. Because one human reference is shown during construction, a natural concern is whether it is
held out from the oracle. It is not: the reference shown to each engine is the first of the eight\evfact{F.atcoder.shown-ref}. Suite construction can optimize against one voter, but one solution cannot form the reference majority. Certification still requires the killed submission to disagree with that majority. Disputes arise when the killed submission agrees with a larger re-judged pool or when the gold, the oracle expectation, is contested. Brute-force solutions adjudicate them. Every kill is re-verified deterministically on a fixed Linux judging harness.

\paragraph{Why legality is part of the certificate.}
\looseness=-1
An audited oracle is insufficient for inputs outside the statement. The final logical link in the chain is a strict per-problem validator. The harness independently checks every certifying input, whether disputed or not. An input enters scoring only if the validator accepts it. For AtCoder, the validator is an agent-produced artifact written by an LLM from the statement alone\evfact{F.atcoder.validator-authorship}. It is blind to the killing inputs it will later judge. This makes the validator independent of the kills it gates. Before use, the AtCoder validator undergoes an instrument-level sanity check, separate from adjudicating constructed findings: it must accept every official input and reject the empty input. If a problem's validator fails this quality gate, every kill row for that problem is scored unmeasured and excluded from the legality count. No such row is scored illegal. Under this gate, legality means validator-passing under the encoded statement constraints. For an AtCoder topological-sort problem that promises a DAG, a cyclic input is rejected and cannot refute a solution \citep{wang-codecontests-2025}. The CF-fresh protocol takes the testlib gate from the reproduced CodeContests+ baseline toolchain and applies the same scoring rule to every arm. \S\ref{sec:why} audits this legality check. Because shipped dataset tests do not validate inputs, legality is an explicit part of the
certificate \citep{wang-codecontests-2025}. Across both experimental regimes, configurations differ, but the certification pattern remains
the same.

\section{Experiment 1: Auditing AtCoder Official Suites}
\label{sec:exp1}

\looseness=-1
Experiment~1 audits two populations against existing AtCoder suites, carrying forward legality
from the \S\ref{sec:method} certificate. For accepted submissions, the certified
\textsc{codex} ledger contains \NAtcoderAcbBuggy{} accepted-but-buggy submissions missed by
the official suites in the audited sample; the five-arm legality- and tolerance-clean floor is
\NAtcoderAcbUnionCleanFloor{}. For rejected submissions, a five-engine panel measures
official-suite logic-bug recovery; its largest engine-specific shortfall is
\NAtcoderRejAgentWorstGap{}pp.

\paragraph{Setup: separating the two audit populations.}
\looseness=-1
The universe is \NAtcoderProblemsTotal{} AtCoder problems with CodeNet human submissions
\citep{puri-codenet-2021}. Differential vetting cross-checks accepted outputs, excluding multi-answer and special-judge
problems; initial scoring uses exact comparison. Per problem, we take the first 200 accepted and 300 rejected compilable C++ submissions. The
accepted sample contains \NAtcoderAuditedAcTotal{} submissions, including all from problems
with fewer than 200. Its ledger uses only the \textsc{codex} arm and full certification chain. Official-suite-rejected submissions form the known-bug panel (REJ). All five engines are
scored separately, never pooled. Engine-specific and random suites use a 50-input cap;
official suites retain baseline sizes. One Linux judging harness scores all. Sample scope is in \S\ref{sec:threats}; full protocols
in \S A.

\paragraph{What survives correction into the ledger.}
\looseness=-1
Within this accepted sample, \textsc{codex} suites expose \NAtcoderAcbBuggy{} verified
accepted-but-buggy submissions missed by the official suites\evclaim{CL.atcoder.acb-count}.
The \NAtcoderAuditedAcTotal{} audited submissions cover \NAtcoderAuditedSharePct\% of these
problems' C++ accepted submissions. Within that sample, the ledger rate is \NAtcoderAcbRate\%. The ledger comprises
\NAtcoderAcbWa{} wrong-answer and \NAtcoderAcbRe{} runtime-error submissions across
\NAtcoderAcbProblems{} problems. Each fails on a legal input under an audited expected output.
The funnel begins with the \textsc{codex}-arm logic-bug exclusives, submissions killed as
logic bugs by its suites but not by official suites (Table~\ref{tab:audit-anatomy}); official
TLE reconciliation and reproducibility filtering leave \NAtcoderAcbRawDet{} deterministic
flags. An \NAtcoderConsensusVoterSample{}-submission sample from the same per-problem accepted pool
cross-checked the reference oracle by voting on each flag. A majority favoring the killed
submission made that oracle the outlier, withdrawing \NAtcoderConsensusFp{} flags. The full
200-submission pool confirmed the \NAtcoderAbcFp{} of these reversals that fall in one problem. Tolerance removed
\NAtcoderTolFp{} floating-point artifacts. Post hoc, independently written per-problem reference solvers re-adjudicated all
\NAtcoderOracleCensusN{} ledger entries in isolated scratch without access to repository oracle
data, using \NAtcoderCensusSolverInputs{} distinct killing inputs; the reviewing model saw neither
expected/got values nor buggy source for wrong-answer rows (\S A). All \NAtcoderAcbRe{}
runtime-error rows received separate static crash attribution, and the re-adjudication disagreed
with \NAtcoderOracleCensusFp{} oracle outputs.

\begin{table}[t]
\centering
{\small
\begin{tabular}{@{}lr@{}}
\toprule
 & submissions \\
\midrule
\textsc{codex}-arm logic-bug exclusives & \NAtcoderOnlyAdgExcl \\
$-$ official TLE catches & \NAtcoderAcbOfficialTleCaught \\
$=$ ledger candidates (no official catch) & \NAtcoderAcbLedgerCandidates \\
\midrule
$-$ nondeterministic / unreproduced & \NAtcoderAcbNondeterministic \\
$=$ raw deterministic kills & \NAtcoderAcbRawDet \\
\midrule
$-$ consensus corrections (\NAtcoderConsensusVoterSample{}-voter; \NAtcoderAbcFp{} full-pool) & \NAtcoderConsensusFp \\
$-$ tolerance false positives & \NAtcoderTolFp \\
\midrule
$=$ verified accepted-but-buggy & \NAtcoderAcbBuggy \\
\quad wrong-answer / runtime-error & \NAtcoderAcbWa{} / \NAtcoderAcbRe \\
\quad problems covered & \NAtcoderAcbProblems{} / \NAtcoderProblemsTotal \\
\bottomrule
\end{tabular}}
\caption{Audit funnel from \textsc{codex}-arm logic-bug exclusives on the accepted pool to the
verified accepted-but-buggy ledger. Official TLE catches reconcile that scoring count;
reproducibility filtering precedes the two artifact corrections.}
\label{tab:audit-anatomy}
\end{table}

\paragraph{The same gates across all five arms.}
\looseness=-1
The arm-general chain runs deterministic re-judge, consensus correction, legality validation,
and tolerance filtering. Within the audited sample, every arm receives the \textsc{codex} ledger's legality and
tolerance gates; they remove no \textsc{codex} rows, so its count remains unchanged. Certified-clean counts are \NAtcoderAcbArmCodexLegal{}
(\textsc{codex}), \NAtcoderAcbArmClaudeLegal{} (\textsc{claude}), \NAtcoderAcbArmAgyLegal{}
(\textsc{agy}), \NAtcoderAcbArmOpencodeLegal{} (\textsc{opencode}), and \NAtcoderAcbArmMiniLegal{}
(\textsc{mini}). Their deduplicated union is a legality- and tolerance-clean floor of
\NAtcoderAcbUnionCleanFloor{} accepted-but-buggy submissions\evclaim{CL.atcoder.acb-union}. The gates drop \NAtcoderAcbUnionGateRejected{} submissions whose every killing input is
illegal and remove \NAtcoderAcbUnionTolFp{} floating-point false positives; because dropped
candidates are not regenerated, the union is a floor. Census confirmation covers
\NAtcoderAcbUnionCleanConfirmed{} of the floor across every arm's certified rows, not only the
\textsc{codex} ledger. A signed, human-adjudicated stratified sample of
\NAtcoderAcbUnionFdrSampleN{} findings had \NAtcoderAcbUnionFdrFp{} overturns: the coverage layer
sampled \NAtcoderAcbUnionFdrCoverageN{} rows within problem with caps and inverse-probability
weighting, and the single risk-layer row was the one left unconfirmed by the census. The coverage
layer's false-discovery rate is bounded at \NAtcoderAcbUnionFdrUpperPct\% (Wilson 95\% upper); the
full protocol is in \S A. The floor comprises the \textsc{codex} ledger's
\NAtcoderAcbBuggy{} submissions plus \NAtcoderAcbUnionNew{} added by the four other arms but
not certified by \textsc{codex}, totaling \NAtcoderAcbUnionCleanFloor{}.

\begin{figure}[t]
\centering
\providecommand{\mainplotwidth}{\columnwidth}
\includegraphics[width=\mainplotwidth]{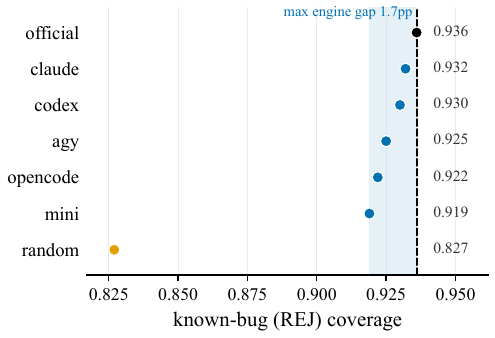}
\caption{Micro-pooled logic-bug coverage (\NAtcoderRejUnionPool{} REJ submissions;
\NAtcoderProblemsTotal{} AtCoder problems; TLE-only excluded; agent and random suites: 50-input
cap; official suites: baseline sizes). Every engine tracks the re-judged official suites closely,
while capped random generation trails.}
\label{fig:parity}
\end{figure}

\paragraph{Known-bug coverage is close, with an official edge.}
\looseness=-1
REJ recovery probes performance on a broad official-known-bug panel. Time-limit verdicts are
excluded from all bug claims (\S A).
Slow accepted solutions' bug status depends on unrecoverable setter intent. Micro-pooled
coverage is \NAtcoderRejAgentCovWorst{}--\NAtcoderRejAgentCovBest{}, versus
\NAtcoderRejOfficialCov{} for re-judged official suites.\footnote{The official rate
\NAtcoderRejOfficialCov{} falls below full recovery of its own REJ pool because of scoring and
harness effects, not missed bugs. Most of the gap is submissions the official inputs catch as
timeouts, which the excl-TLE basis credits to no arm; \S A gives the decomposition and a
sensitivity check.}
The official rate remains higher, but every engine stays within \NAtcoderRejAgentWorstGap{}pp;
agent-cap-matched random generation falls \NAtcoderRejRandomGap{}pp
short\evclaim{CL.atcoder.parity} (Figure~\ref{fig:parity}).
These rates are descriptive; macro-recall, not this pooled rate, underlies the noninferiority test.
Without formal preregistration, a pre-execution plan designated \textsc{codex} and \textsc{claude}
confirmatory, and \NAtcoderNiCiNoninferior{} engines clear the formal test (\S A); the five-engine
result is descriptive, not formal.

\paragraph{Inside the ledger: ruling out a few-template explanation.}
The known-bug comparison tells against official leniency as the ledger's source; its
\NAtcoderAcbBuggy{} verified entries also do not reduce to a few source-level near-clone groups:
mechanical source-token clustering yields \NAtcoderAcbNearclone{} near-clone-distinct clusters
among those findings\evclaim{CL.atcoder.acb-diverse}. The clustering criterion and transcribed
examples are in \S C. Official-test input overlap is audited in \S\ref{sec:threats}.

\paragraph{The official suites remain the stronger single judge.}
\looseness=-1
Sharing a logic-bug basis, the exclusive counts answer opposite cross-pool questions. On REJ,
official suites reject \NAtcoderOnlyOfficialExcl{} submissions missed by \textsc{codex}
suites. On accepted submissions, the converse \textsc{codex}-arm count opens the funnel in
Table~\ref{tab:audit-anatomy}; \NAtcoderAcbBuggy{} survive certification. The counts show
complementary directions, not comparable magnitudes. Expert-built official suites remain the human
baseline and stronger single judge. The certified \textsc{codex} ledger and five-arm floor complement rather than replace official
suites. The next experiment starts where official-suite availability ends.

\section{Experiment 2: Post-Cutoff Head-to-Head on Fresh Codeforces}
\label{sec:exp2}

\looseness=-1
Without access to an official hidden suite, this section compares supplied suites head-to-head. The
agent arm leads every reproduced baseline at every tested input budget.

\paragraph{The supply setting.}
\looseness=-1
The benchmark is \emph{CF-fresh}: \NCfProblems{} Codeforces problems from completed official
rounds harvested under a March 1--June 9, 2026 date rule, intended to postdate the suite-building
engine's declared training-data cutoff: Jan 2026, which precedes the earliest included problem,
dated March 8, 2026\evfact{F.cf.builder-cutoff}. The platform releases statements and sample tests, not full suites. A fixed-seed, content-blind harvest applies date, rating-band, and tag rules,
yielding \NCfHarvestProblems{} problems; rule-based exclusion of \NCfExcludedProblems{}
multi-solution or special-judge cases leaves \NCfProblems{}. Freshness removes the direct
memorization channel; residual controls are in \S\ref{sec:threats}.

\paragraph{Buggy-pool membership and comparison arms.}
\looseness=-1
Among \NCfSolverPoolTotal{} multi-model LLM solver attempts on \NCfProblems{} problems,
\NCfSamplePassing{} compile and pass all public samples. The six arms are the agent arm (one engine, \textsc{claude}) and five reproduced baselines:
official sample tests, cheap-random generation matched to the agent's input budget, the CodeContests dataset's
mutation-based generator protocol \citep{li-competitionlevel-2022}, CodeContests+
\citep{wang-codecontests-2025}, and an EvalPlus-style generator \citep{liu-your-2023}. A sample-passing
solution enters the pool when some arm's legal input exposes it and the three-reference consensus
oracle of \S\ref{sec:method} certifies the failure; the pool is the union of the arms' certified
findings. \S\ref{sec:why} audits the oracle's soundness. The pool contains \NCfNBuggy{} solutions with logic bugs, not timeouts. The official sample arm has zero coverage by
construction because every pool member passes the public samples. It records a builder's pre-generator suite. All arms score only legal inputs under the shared validator.

\begin{figure}[t]
\centering
\providecommand{\mainplotwidth}{\columnwidth}
\includegraphics[width=\mainplotwidth]{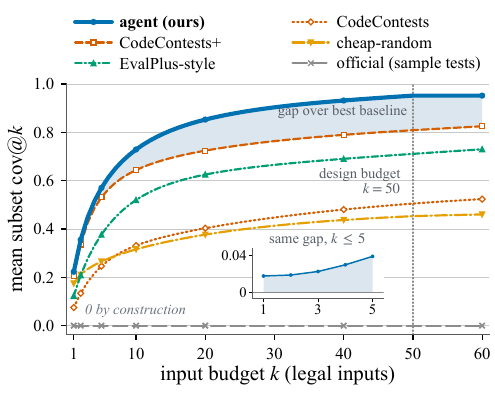}
\caption{CF-fresh subset coverage cov@$k$ (order-free) for all six arms. Coverage averages over
\NCfNBuggy{} adjudicated buggy solutions. The official arm comprises the sample tests, so its zero
coverage is by construction: all pool members pass them. The inset magnifies the gap for $k\,{\leq}\,5$; the dotted line
marks the agent's \NCfAdgFinalPerProblem{}-input design budget, beyond which its full set is used.}
\label{fig:cf-curves}
\end{figure}

\paragraph{Fair comparison basis.}
\looseness=-1
We reproduce the five baselines end-to-end in one kill matrix, paired by problem.
Coverage is cov@$k$, the order-free expected pool fraction killed by a uniform random
$k$-subset of an arm's legal inputs for that problem. An arm with fewer than $k$ legal inputs
contributes its whole set. On the union pool, cov@$k$ gives a relative ordering of the compared arms, not an absolute
rate. Figure~\ref{fig:cf-curves} shows the curves; \S B reports the exact cov@$k$ ladder and per-arm
input counts.

\paragraph{Lead at every input budget.}
\looseness=-1
The agent arm catches more of the CF-fresh buggy pool than every reproduced
baseline\evclaim{CL.cf.adg-beats-baselines} (Figure~\ref{fig:cf-curves}), with the lead
holding at every $k \in [1, 60]$ with no crossover. The minimum gap is \NCfAdgLeadMinGap{} at $k{=}1$. At the
design budget $k{=}50$, the agent finalizes \NCfAdgFinalPerProblem{} inputs on 40 of the
\NCfProblems{} problems (30 on the remaining one). A per-problem mean of \NCfAdgInputsMean{} inputs survives both consensus and legality
gates---the fewest among the generator arms, whose post-top-up means span
\NCfBaselineInputsMin{}--\NCfBaselineInputsMax{} (\S B). At that budget, agent coverage is \NCfAdgCovFifty{} versus \NCfCovFiftyCcplus{} for
CodeContests+, the strongest baseline; equivalently, its miss rate is
\NCfMissRatioCcplusFifty$\times$ the agent's. That budget was fixed by design, not by the curves; we
do not headline cov@60 because the agent's finalization cap is \NCfAdgFinalPerProblem{} inputs.

\paragraph{Budget-dependent separation.}
\looseness=-1
All-budget ordering does not imply per-budget separation. Under the canonical bootstrap
seed, the paired cluster-bootstrap delta against CodeContests+, the strongest baseline, is
\NCfAdgCcplusDeltaFifty{} at the design budget; its 95\% confidence interval
$[\NCfAdgCcplusCiLoFifty,\,\NCfAdgCcplusCiHiFifty]$ excludes zero. All 12 robustness-seed intervals also have positive lower bounds. At the smaller $k{=}20$, the delta is
\NCfAdgCcplusDelta{}, with CI $[\NCfAdgCcplusCiLo,\,\NCfAdgCcplusCiHi]$. That interval crosses zero.
The small-budget lead is directional and under-powered, and we do not claim separation there.
Intervals against CodeContests, the EvalPlus-style generator, and cheap-random exclude zero at both
budgets.

\paragraph{Within-model workflow check.}
\looseness=-1
The coverage lead leaves one alternative explanation: the base model rather than the agentic
workflow. For this check the agent arm is \textsc{mini}, on a separately re-adjudicated subset, not
the \textsc{claude} arm or the main table's pool. We test a single-call arm using the same base
model in four independent reps. No execution result is returned to the model, and buggy-pool
membership is frozen. After adding the arm, we re-execute the kill matrix on the
\NCfBestofnNProblems{} problems containing \NCfBestofnNBuggy{} buggy solutions; \S B tabulates it.
We judge both arms identically. The comparison
varies the wrapper, including tool and iteration affordances and the prompt contract needed to use
them; the matched dimensions and the prompts themselves appear in \S D. At $k{=}50$, \textsc{mini} reaches cov@50 $=$
\NCfBestofnAgentCovFifty{}, versus \NCfBestofnSingleCovFifty{} for the single-call arm, which
misses \NCfBestofnSingleMissRatioFifty$\times$ as many buggy solutions. On the validator-legal
basis, the paired \textsc{mini}-minus-single delta is \NCfBestofnSingleDeltaFifty{}
($[\NCfBestofnSingleCiLoFifty,\,\NCfBestofnSingleCiHiFifty]$) at $k{=}50$. A
deployable four-call 50-slot mixture interleaves the four independent samples by position within
the same input budget, without held-out selection. For this ladder, CodeContests+ is replicated
on the same base model and averaged over four sampling replicates; it is a different pool and
protocol from the CodeContests+ arm above. At $k{=}50$, miss rates are
\NCfBestofnCcplusDsMeanMissFifty{} for CodeContests+, \NCfBestofnSingleMissFifty{} for mean
single-call, \NCfBestofnUnionMissFifty{} for the mixture, and \NCfBestofnAgentMissFifty{} for
\textsc{mini}; per-replicate CodeContests+ coverage varies widely (\S B). The mixture closes
\NCfBestofnMixtureGapClosedFifty\% of the observed single-call--\textsc{mini} gap. In this
replication, CodeContests+ misses \NCfBestofnCcplusDsMissRatioFifty$\times$ as many buggy
solutions as \textsc{mini}. Both paired
contrasts---\textsc{mini} against mean single-call and against CodeContests+---exclude zero at
both budgets; the mixture interval crosses zero. Thus, at the fixed $50$-input deployment budget and
for this four-replicate realization, no-oracle mixing weakens but does not exclude a
sampling-diversity explanation. The held-out, per-problem oracle best-of-$m$ upper bound appears
in \S B. The shortfall is not an inability to construct large cases. Without returned execution
results, a single call receives no signal about whether its outputs pass the validator; validator-legal
rates, maximum-scale input shares, and the zero-legal replicates are reported in \S B. The control
also does not match inference spend (\S\ref{sec:threats}). Within this pairing, the coverage edge
tracks the wrapper as a whole rather than a different base model.

\section{Why the Agent Suites Catch So Many: Artifact Audits and Diversity Correlates}
\label{sec:why}

\looseness=-1
Two separate questions are whether the kills are real in both experiments and what accounts
for Experiment~1's volume and Experiment~2's continuing gains. The evidence supports two
answers. The kills survive direct oracle and legality checks. The volume and continuing gains
are consistent with diversity across the input sets, not one unusually strong probe.

\paragraph{What survives the oracle audits.}
\looseness=-1
The first artifact route is a wrong expected output. On CF-fresh, an independent panel of
sample-passing LLM solutions votes on each gold (the oracle's expected output) for the
\NCfOracleGoldUsable{} agent-arm inputs carrying a usable gold; baseline-arm inputs are
excluded. A case is
contested whenever the largest non-gold vote count exceeds the gold's, a rule met by
\NCfOracleContestedTotal{} cases. The audit resolves all in the oracle's favor, leaving zero
demonstrated wrong golds\evclaim{CL.cf.oracle-sound}. Each is therefore a false alarm: several
LLM solutions share the same wrong answer and out-vote a correct gold from three
system-test-passing human solutions. Brute-force-feasible cases are confirmed directly; the
others on structural grounds. \S B reports the split; the conclusion is confined to that
oracle and those problems. The correction funnel in \S\ref{sec:exp1} establishes AtCoder kill
soundness; the remaining calibration question is whether an
\NAtcoderOracleRefsPerProblem-reference consensus is more machinery than the result needs. It
is conservative by design. When the same kills are re-scored under smaller reference consensuses, a single reference retains
only \NAtcoderOracleKOneRecov{} of the \NAtcoderAcbBuggy{} certified kills, whereas three
references already retain \NAtcoderOracleKThreeRecov{}; the full
\NAtcoderOracleRefsPerProblem-reference consensus is more than
needed\evclaim{CL.atcoder.oracle-frugal}. The sensitivity is in wrong-answer
kills; runtime-error kills are unchanged across the tested reference counts.

\paragraph{Legality checks on the input side.}
\looseness=-1
After the oracle checks, a second apparent-kill route remains: a statement-forbidden input.
Such a failure proves nothing. Here we measure the legality gate of \S\ref{sec:method}.
Independent re-verification finds \NAtcoderKillInputLegal{} of the \NAtcoderAcbBuggy{}
certified AtCoder killing inputs passing strict per-problem statement validators, with zero
illegal and zero unmeasured. By construction, CF-fresh's kill matrix scores legal-only inputs;
the gate is applied on \NCfKillmatrixValApplied{} of \NCfProblems{}
problems\evclaim{CL.method.input-legality}. Each AtCoder agent-produced validator uses only
the statement, never sees killing inputs, and before use must accept every official input and
reject the empty input; all \NAtcoderAcbProblems{} ledger problems pass that quality gate.
Under the legality definition of \S\ref{sec:method} (legality means validator-passing under
the encoded statement constraints), the checks above rule out validator-judged illegal inputs
as the source of these kills, addressing the failure mode in shipped dataset tests
\citep{liu-llmpowered-2024,wang-codecontests-2025}; the validators' own soundness against
malformed inputs is audited separately in \S\ref{sec:threats}.

\paragraph{Saturation locates diversity in the set.}
\looseness=-1
After both artifact checks, validity gives way to volume. The regimes use different proxies
for the same construct, input-set diversity: TCE gives the saturation shape, and AtCoder
trajectories show set construction. The TestCase-Eval (TCE)
known-bug setting traces input-by-input coverage accumulation, a saturation profile the two
main experiments do not expose \citep{yang-can-2025}. It compares the agent input set with
cheap-random generation matched to that input budget. The sweep shows cheap-random front-loading coverage
at normalized cov@1 $\NTceSatOneRandom$ versus the agent at $\NTceSatOneAgent$, then flattening
so that the tail ordering reverses: per-input marginal coverage is
$\NTceMargAgent\pm\NTceMargAgentSe$ for the agent versus $\NTceMargRandom\pm\NTceMargRandomSe$
for cheap-random. The no-feedback control tests whether this tail advantage depends on execution
feedback and gives a comparable per-input marginal coverage of
$\NTceMargAgentNofb\pm\NTceMargAgentNofbSe$; the TCE sweep design, full curves, ratios, and
per-problem tail marginals are in \S B. Computed on the
unrounded tail marginals, the agent has $\sim\NTceMargRatio\times$ higher per-input marginal
coverage\evclaim{CL.tce.diversity}. This is a saturation-shape result, not a measurement of
pairwise input disjointness.

\paragraph{Strategy presence and co-occurrence in the trajectories.}
In the audited AtCoder trajectories from Experiment~1's \textsc{codex} arm, set assembly is
better described as multi-strategy construction than bulk randomness (\S A, \S C).

\paragraph{How the two answers fit the experiments.}
\looseness=-1
The oracle audits and legality certification close the artifact routes; the observed volume is
consistent with diversity across the input sets. Experiment~1's findings are dispersed,
forming \NAtcoderAcbNearclone{} source clusters and \NAtcoderAcbFamilies{} bug-class families
rather than one dominant exploit; the largest single problem contributes only
\NAtcoderAcbTopShare\% of the clusters (\S C). Experiment~2 shows no crossover as suite
size grows (\S\ref{sec:exp2}). At $k{=}1$, cheap-random leads in TCE's human-wrong-submission
pool; CF-fresh's sample-passing-LLM-solution pool has no crossover. We do not offer a
mechanism-level account linking the
composition difference to the observed crossover behavior. The agent sets continue adding
coverage in both pools. The diversity evidence is associative, not causal; the TCE mechanism
estimate does not transfer mechanically beyond the studied population.

\section{Threats to Validity and Limitations}
\label{sec:threats}

\paragraph{Sample scope.}
\looseness=-1
The deterministic sample contains each problem's first 200 compilable submissions
(\NAtcoderAuditedSharePct\% of the accepted pool; \S\ref{sec:exp1}).
The \NAtcoderAcbBuggy{} count remains an in-sample lower bound, not a platform rate; we do not
extrapolate beyond this slice. Matching a future audit to the platform's natural submission
distribution would require CodeNet's full distribution. Compared arms' certified-finding union
bounds the CF-fresh buggy pool; arm-independent pool construction remains open. Rebuilding the
pool without the agent's certified findings leaves it unchanged---every entry is caught by at
least one non-agent arm. The chain assumes
competitive-programming structure---accepted-solution consensus oracle, statement-derived
validator, exact-output judging---and has yet to be extended to specification-only tasks or tasks
without accepted solutions to seed the oracle.

\paragraph{What the legality gate is, and is not.}
\looseness=-1
Validator-passing does not establish semantic ground truth: quality-gated validators
(\S\ref{sec:why}) encode statement-declared constraints, but ambiguity can remain beyond their
reach. Malformed-input gate soundness was audited structurally (trailing-token, truncation, and
non-numeric mutations to official legal inputs: \NAtcoderValidatorFuzzFalseAccepts{} false
accepts) and semantically (format-valid but illegal inputs:
\NAtcoderValidatorSemanticFalseAccepts{} false accepts). A validator-blind, different-family
model constructed both tracks' fuzzing inputs, preventing illegal-input admission via a shared
generator--validator blind spot. Validator misclassification is one-directional: only false
accepts could produce false kills; over-strictness merely under-counts. \S A reports the fuzzing
taxonomy, per-track counts, and an after-the-fact, single-author, non-blind manual
problem-statement check of the then-present AtCoder and CF-fresh validators, finding no
constraint-encoding errors; this attestation does not establish semantic ground truth.

\paragraph{Cost.}
\looseness=-1
Budget matching equalized retained-input counts, not spend; per-problem medians are minutes-scale (\S A).

\paragraph{Contamination.}
\looseness=-1
AtCoder problems predate the engines' training cutoffs, so statement memorization remains possible
in Experiment~1; Experiment~2 uses post-cutoff problems to remove that training-time channel. The
audit concerns the suite, not the model: ``the official suite accepts these buggy submissions''
does not depend on how the killing input was conceived. Experiment~1 measured
agent-generated--official-human-setter input overlap; the five AtCoder engines' highest
non-trivial rate is \NAtcoderESevenOverlapMaxPct\%. Construction occurs in per-run child working
directories; the held-out buggy-submission pool, official test I/O, and CodeNet AC pool remain
physically outside them and are not provided as task inputs.
A post-hoc audit of the recorded shell and message actions from all five engines, checking
filesystem access to held-out materials and official test data, found
\NAtcoderLeakAuditAccessLeaks{} access-level leaks over \NAtcoderLeakAuditScannedRuns{} scanned
agent runs\evfact{F.method.leak-audit}.

\section{Conclusion}
\label{sec:conclusion}

\looseness=-1
Given only a problem statement and one reference solution, off-the-shelf coding agents can audit test-suite adequacy.
\textsc{codex} finds \NAtcoderAcbBuggy{} verified accepted-but-buggy submissions among \NAtcoderAuditedAcTotal{} audited; identically gated, the five-arm union floor reaches \NAtcoderAcbUnionCleanFloor{}, including \NAtcoderAcbUnionNew{} not certified by \textsc{codex}.
Scored separately, all five engines stay within \NAtcoderRejAgentWorstGap{}pp of the re-judged official suites on the logic bugs those suites catch (\S\ref{sec:exp1}).
As a single-engine judge for fresh problems without official suites, it leads every reproduced baseline at all tested input budgets.
Audited oracles underpin both regimes; every certified killing input passes a per-problem legality validator (\S\ref{sec:why}).

Machine-checkable certificates in a re-checkable ledger support a deployable second judge.
Suite adequacy becomes audited rather than assumed on reputation.

\bibliographystyle{plainnat}
\IfFileExists{main.bib}
  {\bibliography{main}}
  {\bibliography{../main}}

\clearpage
\appendix
\section*{Appendix}
\setcounter{figure}{0}
\setcounter{table}{0}
\setcounter{equation}{0}
\renewcommand{\thefigure}{\thesection.\arabic{figure}}
\renewcommand{\thetable}{\thesection.\arabic{table}}
\renewcommand{\theequation}{\thesection.\arabic{equation}}
\renewcommand{\theHfigure}{appendix.\thesection.\arabic{figure}}
\renewcommand{\theHtable}{appendix.\thesection.\arabic{table}}
\renewcommand{\theHequation}{appendix.\thesection.\arabic{equation}}

\providecommand{\mainpapername}{main paper}
\providecommand{\mainpapersection}[1]{main paper \S#1}

\section{Audit Protocol Details (AtCoder)}
\label{app:audit}

This appendix gives the full protocol behind Experiment~1 (\mainpapersection{4}): the submission
sampling and the two audit populations, how the consensus oracle is built and audited, the
judging environment, and the per-entry schema of the released ledger. Numbers shared with the main text re-use the same macros; per-engine detail, costs, the
full self-gap decomposition, and some protocol parameters are reported only here.

\paragraph{Problem selection.}
The study problems come from a \NAtcoderSelectionCandidates-problem candidate pool (AtCoder ABC
D/E/F tasks in the mid-difficulty band, present in both the AtCoder mirror and CodeNet \citep{puri-codenet-2021}), narrowed
to the \NAtcoderSelectionEligible{} with a usable reference pool and unambiguous exact-match
verdicts. Multi-answer and special-judge tasks are removed at selection time, so exact output
comparison is the intended verdict basis throughout.

\paragraph{Submission sampling and the two audit populations.}
The universe is \NAtcoderProblemsTotal{} problems with CodeNet human submissions
\citep{puri-codenet-2021}. Per problem, we take the first 200 accepted and the first 300 rejected
compilable C++ submissions. The accepted sample contains \NAtcoderAuditedAcTotal{} submissions,
includes every accepted submission from problems with fewer than 200, and covers
\NAtcoderAuditedSharePct\% of these problems' C++ accepted submissions. Accepted-but-buggy
findings come from this pool, whose ledger uses only the \textsc{codex} arm under the full
certification chain. Officially rejected submissions form the separate known-bug panel (REJ), scored on the pool
defined below rather than on the whole sample; the five engines are scored separately and never
pooled. Engine-specific and random suites use a
50-input cap. Official suites retain their baseline sizes.

\paragraph{Consensus oracle and its audit.}
The oracle draws on \NAtcoderRefCount{} accepted human reference solutions across the
\NAtcoderProblemsTotal{} problems, \NAtcoderOracleRefsPerProblem{} per problem, all
source-level distinct after near-clone deduplication (\NAtcoderRefNearcloneClusters{} clusters).
The expected output for an input is the value the references agree on; a submission is killed
when its output disagrees with that majority, or when it exits with a runtime error rather than
an output. Two properties are audited. Conservatism: on the
\NAtcoderAcbBuggy{} certified kills, a single reference recovers only \NAtcoderOracleKOneRecov{}
of them, while three already recover \NAtcoderOracleKThreeRecov{} and five and eight both hold at
\NAtcoderOracleKFiveRecov{}. A lone reference is too few; eight is more than needed.
Soundness: the correction funnel appears in \mainpapersection{4} and
its audit-anatomy table. In the single-problem \texttt{abc164\_e} concentration, all
\NAtcoderAbcFp{} retractions were checked without sampling against the full 200-submission pool.
Disputes too large for pool re-judging use brute-force adjudication.

\paragraph{Census re-adjudication.}
After the openly-corrected pipeline closed, every one of the \NAtcoderOracleCensusN{} ledger
findings was independently re-adjudicated.
Independently written per-problem reference solvers, validated on the official samples before use,
ran on \NAtcoderCensusSolverInputs{} distinct killing inputs for wrong-answer submissions.
They ran in isolated scratch without access to repository
oracle data. Their prompts omitted expected and got values, bug class, and DET class (the
candidate's stability on its killing input across three re-runs on the Linux judging host:
\texttt{DET\_WA} for the same wrong normalized output throughout, \texttt{DET\_RE} for a runtime
error throughout, \texttt{NONDET} for instability, and \texttt{NOT\_REPRODUCED} if no killing
input reproduces the kill; only \texttt{DET\_WA} and \texttt{DET\_RE} enter the ledger).
Wrong-answer
review saw the statement and killing inputs but not buggy source. Runtime-error review saw source
only for crash attribution, with no expected output.
The authors compared outputs; the reviewing model did not see the result.
The 16 rows whose outputs exceed the ledger's
80-character display window were resolved by full-output comparison against re-run buggy
binaries. All \NAtcoderAcbRe{} runtime-error submissions received static crash attribution
(mechanism and trigger named, input legality confirmed). The solvers disagreed with
\NAtcoderOracleCensusFp{} oracle outputs and had zero conflicts with the earlier pre-registered
stratified sample over the \textsc{codex} arm's pre-audit deterministic ledger: its coverage layer
was a per-problem random draw capped at three per problem, with 0/163 adjudicated rows overturned,
yielding an independent coverage-layer FDR bound of \NAtcoderOracleSampleFdrUpperPct\%
(Wilson 95\% upper); its risk layer exhaustively audited the multi-solution risk classes,
tie-breaking and precision, in which more than one output can legitimately be correct, with 0/56
overturned. The protocol tests shared-error risk rather than assuming it away: on
\texttt{abc145\_e}, the reference solver shared the buggy submissions' misreading of the statement
and was refuted by a dual-route brute-force witness. Disputed problems route to
brute-force adjudication, and rows retracted by the earlier 200-pool and tolerance audits
are counted separately, the upstream correction mechanisms working as designed.

\paragraph{Judging environment.}
Every authoritative kill is re-judged deterministically on a single Linux host; numbers observed
on other platforms during development never enter a claim. The harness enforces each problem's own
time limit plus a 0.5\,s scheduling allowance, and it removes the default stack limit, matching
judge semantics so that deep-recursion accepted solutions are not spuriously killed. Verdicts are
bucketed by type, and time-limit-exceeded catches are held in a separate bucket excluded from
every bug claim in the paper (the excl-TLE basis).

\paragraph{Validator soundness audit.}
The per-problem legality validators (\mainpapersection{7}) are audited against malformed inputs on two
fronts, both built by a validator-blind different-family model so a shared blind spot cannot pass.
Structurally, across \NAtcoderValidatorFuzzProblems{} problems, \NAtcoderValidatorFuzzInputs{}
structurally illegal mutations of official inputs (trailing tokens, truncation, non-numeric fields)
drew \NAtcoderValidatorFuzzFalseAccepts{} false accepts. Semantically, across
\NAtcoderValidatorSemanticProblems{} problems, \NAtcoderValidatorSemanticInputs{}
format-valid-but-illegal inputs (a duplicate in a permutation, a cycle in a tree, a multi-edge in
a simple graph, a broken connectivity or distinctness constraint) drew
\NAtcoderValidatorSemanticFalseAccepts{} false accepts. The gate rejects the malformations it
exists to catch, and its error is one-directional: only a false accept could buy a false kill,
while over-strictness merely under-counts. Complementing these machine audits, an author manually
reviewed every per-problem validator present when the review was recorded---the AtCoder
agent-written validators and the
CF-fresh testlib validators---against its problem statement, checking whether each
statement-declared input constraint was encoded; the review found no constraint-encoding
errors\evfact{F.method.validator-manual-audit}. This single-author, non-blind review was recorded
after the fact and is an attestation, not an establishment of semantic ground truth.

\paragraph{Cost.}
Budget matching equalized retained input counts per problem, not spend. Measured per-problem
medians on the AtCoder panel are \NAtcoderCostOutKtokCodex{}k output tokens /
\NAtcoderCostWallMinCodex{} min (\textsc{codex})\footnote{The five auditing engines are defined in
\mainpapersection{3}; each pairs an off-the-shelf coding-agent CLI with a base model:
\textsc{codex} (OpenAI Codex CLI~\citep{openai-codex-2025}, GPT-5.4~\citep{openai-gpt54-2026}),
\textsc{claude} (Claude Code~\citep{anthropic-claudecode-2025}, Claude Opus
4.8~\citep{anthropic-opus48-2026}), \textsc{agy} (Antigravity CLI~\citep{google-antigravity-2025},
Gemini 3.1 Pro~\citep{google-gemini31-2026}), \textsc{opencode} (OpenCode
CLI~\citep{opencode-2025}, DeepSeek V4 Pro~\citep{deepseek-v4-2026}), and \textsc{mini}
(mini-SWE-agent~\citep{yang-sweagent-2024} on the same DeepSeek V4 Pro). Reasoning-effort settings
are xhigh for \textsc{codex} and \textsc{claude}, high for \textsc{agy}, and thinking-high for
\textsc{opencode} and \textsc{mini}.}, \NAtcoderCostOutKtokClaude{}k /
\NAtcoderCostWallMinClaude{} min (\textsc{claude}), \NAtcoderCostOutKtokOpencode{}k /
\NAtcoderCostWallMinOpencode{} min (\textsc{opencode}), and \NAtcoderCostWallMinMini{} min
(\textsc{mini}). The CF-fresh agent arm spends \NCfCostOutKtok{}k tokens and \NCfCostWallMin{} min
over \NCfCostTurns{} turns per problem.
\textsc{agy} exposes no token or call accounting and no usable wall-clock, because its CLI records
a narration-only transcript with no structured events; the raw spawn-to-last-write span includes
requeue and watchdog waiting, so it is not registered as a generation wall-clock. \textsc{mini}
exposes call counts but not token accounting. The script and reproduced-suite arms carry only
reproduction-side cost.

\paragraph{Per-engine known-bug coverage detail.}
Table~\ref{tab:parity-detail} expands the known-bug guard of \mainpapersection{4} to the five engines
individually. The excl-TLE official-suite-rejected (REJ) pool contains the \NAtcoderRejUnionPool{} historical
submissions that at least one arm's re-judged suite kills through a wrong-answer or runtime-error
verdict; submissions no arm kills that way stay outside the pool, and within it every arm is
credited only for its own wrong-answer and runtime-error catches. On this pool, official coverage
\NAtcoderRejOfficialCov{} leaves a self-gap of \NAtcoderRejSelfgapRows{} submissions
(\NAtcoderRejSelfgapPct\% of the pool). Of this gap, \NAtcoderRejSelfgapTleShare\% is TLE masking:
the official inputs still kill these submissions, but only with a time-limit-exceeded verdict,
which the excl-TLE basis credits for no arm. The remaining \NAtcoderRejSelfgapDrift{} submissions
(\NAtcoderRejSelfgapDriftPct\% of the pool) are verdict drift: under the judging environment
above, the official inputs no longer kill them at all, while at least one other arm's suite does.
Dropping the drift submissions raises official coverage and widens the largest engine-to-official
gap from \NAtcoderRejAgentWorstGap{}pp on the full pool to \NAtcoderRejParityGapReproPp{}pp;
official coverage retains a small edge under both treatments.
Noninferiority is tested per engine on a per-problem macro-recall basis against the
re-judged official suites, with parameters fixed in advance (margin $\Delta{=}0.02$,
$B{=}10{,}000$ bootstrap resamples, seed 20260619; no formal preregistration). An engine is
noninferior at its single-engine 95\% interval when the interval's lower bound stays above
$-\Delta$. The \NAtcoderNiCiNoninferior{} confirmatory engines (\textsc{codex}, \textsc{claude})
clear this bar. Given the internal designation, Holm pass counts are
\NAtcoderNiHolmPassConfirmatory{} for the confirmatory pair and \NAtcoderNiHolmPass{} across five
engines. Random generation matched to the agent's input budget (\NAtcoderRejRandomCov{}
coverage) is well outside it.

\paragraph{Input legality on the known-bug panel.}
The panel scores every arm's inputs \emph{as generated}: unlike the accepted-pool chain, it applies
no legality gate at scoring time. Re-running the same per-problem validators after the fact over
every panel input covers the \NAtcoderRejLegalityProblems{} of \NAtcoderProblemsTotal{} panel
problems that carry a validator. The per-engine share of illegal inputs runs from
\NAtcoderRejLegalityMinPct\% to \NAtcoderRejLegalityMaxPct\%, and the random arm sits at
\NAtcoderRejLegalityRandomPct\% --- inside the engine range, so neither side of the
engine-versus-random contrast is systematically favored. These are input-level shares, and they
bound the exposure rather than correct the coverage rates: a submission is usually killed by
several inputs, so a kill survives a legal-only basis whenever one legal input kills it.

\begin{table*}[t]
\centering
\setlength{\tabcolsep}{4.5pt}
\begin{tabular}{lccc}
\toprule
engine & micro-pooled cov. & macro-recall $\Delta$ & 95\% CI \\
\midrule
\textsc{codex}    & \NAtcoderRejCovCodex    & \NAtcoderNiDeltaCodex    & $[\NAtcoderNiDeltaLoCodex,\,\NAtcoderNiDeltaHiCodex]$ \\
\textsc{claude}   & \NAtcoderRejCovClaude   & \NAtcoderNiDeltaClaude   & $[\NAtcoderNiDeltaLoClaude,\,\NAtcoderNiDeltaHiClaude]$ \\
\textsc{agy}      & \NAtcoderRejCovAgy      & \NAtcoderNiDeltaAgy      & $[\NAtcoderNiDeltaLoAgy,\,\NAtcoderNiDeltaHiAgy]$ \\
\textsc{opencode} & \NAtcoderRejCovOpencode & \NAtcoderNiDeltaOpencode & $[\NAtcoderNiDeltaLoOpencode,\,\NAtcoderNiDeltaHiOpencode]$ \\
\textsc{mini}     & \NAtcoderRejCovMini     & \NAtcoderNiDeltaMini     & $[\NAtcoderNiDeltaLoMini,\,\NAtcoderNiDeltaHiMini]$ \\
\midrule
random            & \NAtcoderRejRandomCov   & --- & --- \\
official          & \NAtcoderRejOfficialCov & --- & --- \\
\bottomrule
\end{tabular}
\caption{Per-engine known-bug (REJ, excl-TLE) results on the AtCoder panel. The two numeric
columns are different estimands and \emph{do not subtract}: coverage is micro-pooled over
submissions, while the noninferiority delta against the re-judged official suites is computed on
a per-problem macro-recall basis. Noninferior at the single-engine interval when the lower bound
exceeds $-\Delta{=}-0.02$. The full per-problem verified accepted-but-buggy distribution over the
\NAtcoderAcbProblems{} problems ships as a CSV with the released artifacts.}
\label{tab:parity-detail}
\end{table*}

\paragraph{Certified per-arm contribution and isolation.}
Separate from the REJ coverage comparison, the arm-general certification chain yields
certified-clean counts of \NAtcoderAcbArmCodexLegal{} (\textsc{codex}), \NAtcoderAcbArmClaudeLegal{}
(\textsc{claude}), \NAtcoderAcbArmAgyLegal{} (\textsc{agy}), \NAtcoderAcbArmOpencodeLegal{}
(\textsc{opencode}), and \NAtcoderAcbArmMiniLegal{} (\textsc{mini}). Their
submission-deduplicated, legality- and tolerance-clean union is a floor of
\NAtcoderAcbUnionCleanFloor{} accepted-but-buggy submissions: \NAtcoderAcbUnionGateRejected{}
candidate submissions whose every killing input was illegal were dropped as invalid findings
rather than regenerated, and \NAtcoderAcbUnionTolFp{} floating-point false positives were
removed. The census protocol above ran on each arm's certified rows, not only on the \textsc{codex} ledger,
and \NAtcoderAcbUnionCleanConfirmed{} of the floor are confirmed: for each of them, at least one arm
that supplies a legal killing input returns a confirming verdict. A separate
stratified sample of \NAtcoderAcbUnionFdrSampleN{} floor findings comprises a coverage layer of
\NAtcoderAcbUnionFdrCoverageN{} rows drawn at random within each problem, capped per problem and
inverse-probability weighted so no single problem dominates the estimate. Its risk layer holds
every floor member outside that confirmed core; its sole row is the one the census left
unconfirmed, because the census's independently written solver produced no usable output for it.
An author adjudicated all sampled rows directly: the unweighted coverage-layer overturn count was
\NAtcoderAcbUnionFdrFp{}, and the risk-layer row was upheld as a legal input with a wrong output. The signed report ships
with the released artifacts, and the coverage layer bounds the false-discovery rate at
\NAtcoderAcbUnionFdrUpperPct\% (Wilson 95\% upper). Of this floor,
\NAtcoderAcbUnionNew{} are certified only by the four non-\textsc{codex} arms, while
\NAtcoderAcbUnionCodexExclusive{} are certified by \textsc{codex} alone. As a raw
pre-certification panel view, the \NAtcoderAcPanelPool{}-row denominator comprises re-judged
accepted submissions with a wrong-answer or runtime-error kill, excluding TLE, from at least one
of the five engine suites, the official suites under re-judge, or random generation. The five
engine suites jointly kill a fraction \NAtcoderAcUnionCov{} of this uncertified panel, while
\NAtcoderAcOnlyOfficialUnion{} rows are killed only by the official suites under re-judge; this
panel fraction is not comparable to the REJ coverage rates. The
\textsc{opencode}/\textsc{mini} pair shares a base model but differs in harness, so their
certified-count gap isolates the generation-side effect of the agent harness from the
underlying model.

\paragraph{Trajectory strategy labels.}
\looseness=-1
The English labels translate the annotation codebook. \emph{Small-case enumeration} systematically
enumerates all small cases, including $n{=}1$, $n{=}2$, single-element, and other trivial inputs.
\emph{Structural / extreme families} exercise named extremes such as maximum-$N$, boundary,
overflow, all-zero or all-negative values, repeated values, degenerate star, chain, tree, or grid
shapes, and worst-case stress. \emph{Semantic-adversarial construction} targets a
problem-specific logical trap in the statement semantics or an algorithmic assumption, rather than
only a structural extreme. \emph{Generator sweep} uses random or seeded batch generation.
\emph{Self-built wrong panel} uses agent-written incorrect solutions as probes. \emph{Paired
differential mining} runs a wrong solution against the reference to mine an input on which their
outputs differ. An LLM assigns these labels from trajectory digests; a same-model blind pass-2
re-label measures test--retest reliability, as reported in Table~\ref{tab:trajectory}.

\paragraph{Ledger schema.}
Each of the \NAtcoderAcbBuggy{} verified entries records the submission fingerprint and its
official verdict, the certifying killing input with its per-problem legality certificate, the
reference pool that adjudicated the kill, and the generation channel it is attributed to.
The ledger covers \NAtcoderAcbProblems{} of \NAtcoderProblemsTotal{} problems and splits into
\NAtcoderAcbWa{} wrong-answer and \NAtcoderAcbRe{} runtime-error findings; the full per-problem
distribution is released as a machine-checkable CSV.

\section{Reproduction Details (CF-fresh and TCE)}
\label{app:repro}

This appendix documents the comparisons behind Experiment~2 (\mainpapersection{5}) and the
mechanism analysis (\mainpapersection{6}): the CF-fresh kill matrix and its baselines, the within-model
single-call control (on its rejudged matrix), and the TestCase-%
Eval (TCE) saturation sweep.

\paragraph{CF-fresh kill matrix.}
The benchmark is \NCfNBuggy{} sample-passing LLM-written solutions (excl-TLE) over \NCfProblems{}
post-cutoff Codeforces problems (finished, non-gym contest rounds harvested under a
March~1--June~9, 2026 window, with included problems spanning 2026-03-08 to 2026-06-07). The
CF-fresh premise is anchored to the suite-building engine: its vendor publishes a declared Jan 2026
training-data cutoff, which precedes the earliest included problem\evfact{F.cf.builder-cutoff}. A
per-engine cutoff table is not available, as Google, DeepSeek, and Alibaba publish no declared
cutoffs for the other models used. The solutions are adjudicated buggy
by the three-reference consensus oracle of \mainpapersection{3} (its soundness audit is
\mainpapersection{6}). Post-cutoff selection removes the direct model-side memorization channel. Harvesting uses two
content-blind passes, a rating-stratified sample and a newest-first full take; interactive or
unrated problems never enter either pass. Of the \NCfHarvestProblems{} harvested problems,
\NCfExcludedProblems{} multi-solution or special-judge cases are excluded post-harvest by rule,
leaving \NCfProblems{}; the released data lists their identifiers and reasons. Of the \NCfSolverPoolTotal{} attempts in the
multi-model solver population (\NCfSolverPoolSingle{} single-shot LLM solutions and
\NCfSolverPoolAgent{} agent-written ones across the source-model ladder), \NCfSamplePassing{}
compile and pass all public samples; the buggy pool keeps the \NCfNBuggy{} that fail the
consensus oracle. Six arms are
scored on one kill matrix, paired per problem, inputs-legal-only under a single \texttt{testlib}\footnote{\texttt{testlib.h}, M.~Mirzayanov,
\url{https://github.com/MikeMirzayanov/testlib}.}
validator gate from the CodeContests+ reproduction toolchain \citep{wang-codecontests-2025}, applied to all \NCfProblems{} problems. The
arms are the agent arm (a single \textsc{claude} engine, no-feedback configuration), the
official sample tests, cheap-random generation matched to the agent's input budget, the mutation-based generator
protocol provided with the CodeContests dataset \citep{li-competitionlevel-2022}, CodeContests+, and an EvalPlus-style generator
\citep{liu-your-2023}. Coverage is cov@$k$, the expected
fraction of the buggy pool killed by a uniform random $k$-subset of an arm's legal inputs
(order-free). An arm with fewer than $k$ legal inputs contributes its whole set.
Table~\ref{tab:cf-full} reports cov@$k$ at eight budgets and gives per-arm input counts. The
comparison was evaluated at every integer budget $k=1,\ldots,60$ and showed no crossover; its
minimum gap was \NCfAdgLeadMinGap{} at $k{=}1$.

\paragraph{Baseline configurations.}
We summarize the concrete implementations, generation procedures, and filtering rules of all
CF-fresh arms in Table~\ref{tab:cffresh-config} before detailing their deviations from the original
baseline descriptions.

\providecommand{\appendixwidetablepadding}{}
\begin{table*}[t]
\centering
\appendixwidetablepadding
{\footnotesize
\begin{tabular}{lllcl}
\toprule
arm & script & LLM calls per problem & target & generation-time filter \\
\midrule
official           & statement samples            & none                          & --- & --- \\
cheap-random       & \texttt{cf\_gen.py}          & 1, up to 2 retries            & 50 & generator self-check, then consensus \\
CodeContests       & \texttt{cf\_codecontests.py} & none                          & 60 & consensus \\
CodeContests+      & \texttt{cf\_ccplus.py}       & 1, up to 2 retries (feedback) & 60 & own validator, then consensus \\
EvalPlus-style     & \texttt{cf\_evalplus.py}     & 1, for seeds                  & 60 & consensus \\
agent (ours)       & agent harness                & agentic loop                  & 50 & reference runs successfully \\
\bottomrule
\end{tabular}}
\caption{Configuration of the CF-fresh input-generation arms used in the kill-matrix comparison.
Scripts are released under \texttt{code/scripts/cf\_fresh/}; every arm that invokes an LLM uses the
same base model, \texttt{claude-4.8-opus}, as the agent arm, holding model capability fixed while
varying the input-generation procedure. The implementations do not exactly match the original
baseline descriptions; the deviations are detailed below.}
\label{tab:cffresh-config}
\end{table*}

CodeContests originally mutated both public and private tests and used consensus over 30 accepted
solutions. For CF-fresh, only public samples are available for mutation because these tasks provide
no private tests; we instead use consensus over three human accepted solutions. Because the original
mutator source was not released, we reconstructed its operator-selection rates and mutation steps
from the paper description. EvalPlus originally used Python runtime type dispatch for type-aware
mutation; stdin exposes no type information, so we classify tokens as integers or alphabetic strings,
apply type-aware mutation and boundary injection, and do not extract a complete input grammar. For
CodeContests+, the original model was not disclosed, so we use the unified base model; we also do
not reproduce the checker layer based on eight \texttt{testlib} checkers for multi-solution tasks.
The cheap-random arm does not reproduce a published baseline; it serves as a floor using a
seed-reading generator program.

\paragraph{Equal legal-input budget.}
We top up the CodeContests and EvalPlus-style arms, and the cheap-random arm when its generation
program supports reuse, before scoring the CF-fresh kill matrix. For each applicable arm, we
continue its own generation procedure and apply validator-first filtering plus three-reference
consensus checks until the legal-input budget reaches \NCfTopupTarget{} inputs. The agent arm does
not participate in this top-up and remains at its original generation budget. On the
\NCfTopupProblems{} of \NCfProblems{} problems with available native-output records, the
pre-top-up per-problem legal-input medians are \NCfTopupNativeCc{} for CodeContests,
\NCfTopupNativeEp{} for EvalPlus-style, and \NCfTopupNativeRandom{} for cheap-random. These
medians report the native outputs before completion to the common legal-input target. On some
problems, none of the applicable generators yields a legal input, so the corresponding arm--problem
entries remain zero. Table~\ref{tab:cf-full} averages inputs equally over all problems, including
those entries, and therefore reports per-problem means below the completion target.

\begin{table*}[t]
\centering
\appendixwidetablepadding
\begin{tabular}{lccccccccr}
\toprule
arm & cov@1 & cov@2 & cov@5 & cov@10 & cov@20 & cov@40 & cov@50 & cov@60 & inputs \\
\midrule
agent (ours)       & \textbf{\NCfAdgCovOne}   & \textbf{\NCfAdgCovTwo}   & \textbf{\NCfAdgCovFive}   & \textbf{\NCfAdgCovTen}   & \textbf{\NCfAdgCovTwenty} & \textbf{\NCfAdgCovForty} & \textbf{\NCfAdgCovFifty} & \textbf{\NCfAdgCovSixty} & \NCfAdgInputsMean \\
CodeContests+      & \NCfCovOneCcplus         & \NCfCovTwoCcplus         & \NCfCovFiveCcplus         & \NCfCovTenCcplus         & \NCfCovCcplus & \NCfCovFortyCcplus & \NCfCovFiftyCcplus & \NCfCovSixtyCcplus & \NCfInputsCcplus \\
EvalPlus-style     & \NCfCovOneEvalplus       & \NCfCovTwoEvalplus       & \NCfCovFiveEvalplus       & \NCfCovTenEvalplus       & \NCfCovEvalplus & \NCfCovFortyEvalplus & \NCfCovFiftyEvalplus & \NCfCovSixtyEvalplus & \NCfInputsEvalplus \\
CodeContests       & \NCfCovOneCodecontests   & \NCfCovTwoCodecontests   & \NCfCovFiveCodecontests   & \NCfCovTenCodecontests   & \NCfCovCodecontests & \NCfCovFortyCodecontests & \NCfCovFiftyCodecontests & \NCfCovSixtyCodecontests & \NCfInputsCodecontests \\
cheap-random       & \NCfCovOneRandom         & \NCfCovTwoRandom         & \NCfCovFiveRandom         & \NCfCovTenRandom         & \NCfCovRandom & \NCfCovFortyRandom & \NCfCovFiftyRandom & \NCfCovSixtyRandom & \NCfInputsRandom \\
official (samples) & \NCfCovOneOfficial       & \NCfCovTwoOfficial       & \NCfCovFiveOfficial       & \NCfCovTenOfficial       & \NCfCovOfficial & \NCfCovFortyOfficial & \NCfCovFiftyOfficial & \NCfCovSixtyOfficial & \NCfInputsOfficial \\
\bottomrule
\end{tabular}
\caption{CF-fresh kill matrix, cov@$k$ at eight budgets over the \NCfNBuggy{} sample-passing
adjudicated-buggy solutions (legal-only, excl-TLE, paired per problem), with mean legal inputs
per problem (an equal-weight mean over all \NCfProblems{} problems, counting an arm--problem pair
as zero when the consensus gate filters all of its inputs). The agent has the highest coverage in
every cov@$k$ column without having the most inputs. Best per column in bold.}
\label{tab:cf-full}
\end{table*}

Table~\ref{tab:cf-single} reports the deployable within-model controls on their rejudged
basis, including the mean single-call arm, the position-interleaved mixture, and the
same-base-model CodeContests+ replication. On this base model,
\NCfBestofnCcplusDsGvInconsistent{} of \NCfBestofnNProblems{} problems produce a generator whose
output its own generated validator rejects, against
\NCfBestofnCcplusGvInconsistentOpus{} on the opus-based replication; a replicate that draws such a
generator keeps no inputs for that problem.
Table~\ref{tab:cf-oracle-bound} then reports a retrospective held-out-pool upper bound as
a diagnostic rather than as an entry in the baseline ranking.

\begin{table*}[t]
\centering
{\footnotesize
\setlength{\tabcolsep}{3pt}
\begin{tabular}{lcc}
\toprule
configuration & cov@20 & cov@50 \\
\midrule
\textsc{mini} (agent)     & \NCfBestofnAgentCovTwenty & \NCfBestofnAgentCovFifty \\
mean single-call          & \NCfBestofnSingleCovTwenty & \NCfBestofnSingleCovFifty \\
four-call 50-slot mixture & \NCfBestofnUnionCovTwenty & \NCfBestofnUnionCovFifty \\
CodeContests+ (four-rep mean) & \NCfBestofnCcplusDsMeanCovTwenty & \NCfBestofnCcplusDsMeanCovFifty \\
\midrule
$\Delta$ vs single (95\% CI) & \NCfBestofnSingleDeltaTwenty{} [\NCfBestofnSingleCiLoTwenty,
\NCfBestofnSingleCiHiTwenty] & \NCfBestofnSingleDeltaFifty{} [\NCfBestofnSingleCiLoFifty,
\NCfBestofnSingleCiHiFifty] \\
$\Delta$ vs mixture (95\% CI) & \NCfBestofnUnionDeltaTwenty{} [\NCfBestofnUnionCiLoTwenty,
\NCfBestofnUnionCiHiTwenty] & \NCfBestofnUnionDeltaFifty{} [\NCfBestofnUnionCiLoFifty,
\NCfBestofnUnionCiHiFifty] \\
$\Delta$ vs CodeContests+ (95\% CI) & \NCfBestofnCcplusDsDeltaTwenty{} [\NCfBestofnCcplusDsCiLoTwenty,
\NCfBestofnCcplusDsCiHiTwenty] & \NCfBestofnCcplusDsDeltaFifty{} [\NCfBestofnCcplusDsCiLoFifty,
\NCfBestofnCcplusDsCiHiFifty] \\
\bottomrule
\end{tabular}}
\caption{Deployable within-model controls on the rejudged basis of \NCfBestofnNProblems{}
problems and \NCfBestofnNBuggy{} buggy solutions. The agent row is \textsc{mini}, rather than the
\textsc{claude} arm in Table~\ref{tab:cf-full}; that table uses a different buggy-solution pool, so
cross-table subtraction does not define a matched contrast. The mean single-call row averages four
independent reps. The mixture interleaves those reps by position within one $50$-input suite, with
no held-out selection. The CodeContests+ row is a same-base-model replication, averaged
over four sampling replicates and construction-matched to the mean single-call row; it is not
interchangeable with the single-sample, opus-based CodeContests+ arm of Table~\ref{tab:cf-full}
(different base model, sampling protocol, and pool). Its per-replicate cov@50 values are
\NCfBestofnCcplusDsRepOneCovFifty{}, \NCfBestofnCcplusDsRepTwoCovFifty{},
\NCfBestofnCcplusDsRepThreeCovFifty{}, and \NCfBestofnCcplusDsRepFourCovFifty{}; a single sample
of this protocol is not a stable estimate. At both budgets, the \textsc{mini}--single-call and
\textsc{mini}--CodeContests+ 95\% confidence intervals exclude zero, whereas the
\textsc{mini}--mixture interval crosses zero. Values in different rows are rounded
independently.}
\label{tab:cf-single}
\end{table*}

\begin{table*}[t]
\centering
{\footnotesize
\setlength{\tabcolsep}{3pt}
\begin{tabular}{lcc}
\toprule
non-deployable bound & cov@1 & cov@50 \\
\midrule
\textsc{mini} (reference) & \NCfBestofnAgentCovOne & \NCfBestofnAgentCovFifty \\
oracle best-of-2          & ---                    & \NCfBestofnOracleTwoCovFifty \\
oracle best-of-3          & ---                    & \NCfBestofnOracleThreeCovFifty \\
oracle best-of-4          & \NCfBestofnOracleFourCovOne & \NCfBestofnOracleFourCovFifty \\
\midrule
$\Delta$ vs best-of-4 (95\% CI) & \NCfBestofnOracleFourDeltaOne{} [\NCfBestofnOracleFourCiLoOne,
\NCfBestofnOracleFourCiHiOne] & \NCfBestofnOracleFourDeltaFifty{} [\NCfBestofnOracleFourCiLoFifty,
\NCfBestofnOracleFourCiHiFifty] \\
\bottomrule
\end{tabular}}
\caption{Non-deployable retrospective oracle upper bounds, excluded from baseline ranking. For each
problem, the oracle selects the best of the four single-call reps using the held-out buggy pool;
that post hoc selection is unavailable at deployment. At cov@1, best-of-4 ranks above
\textsc{mini}: the paired \textsc{mini} minus best-of-4 difference is negative, and its 95\%
confidence interval excludes zero. Retrospective selection has its greatest leverage at this suite
size because single-draw variance is greatest there; this variance account does not alter the
ordering reversal. At cov@50 the paired interval crosses zero, so the bound and \textsc{mini} are
not separated. The cov@1 column is included as a diagnostic and is not the paper's operating
point.}
\label{tab:cf-oracle-bound}
\end{table*}

Within the same controls, the single-call shortfall is not an inability to construct large cases:
maximum-scale inputs comprise \NCfTypemixSingleMaxsizeRateMin--\NCfTypemixSingleMaxsizeRateMax{}
of the single-call arm's legal inputs across reps, versus \NCfTypemixAgentMaxsizeRate{} for
\textsc{mini}, the smallest share of any arm here. Compilation failure, violation of a stated input
constraint, and structurally malformed output are alike absorbed by the consensus oracle. Across
the four replicates, validator-legal rates span \NCfBestofnSingleLegalRateMin{}--%
\NCfBestofnSingleLegalRateMax{}, versus \NCfBestofnAgentLegalRate{} for \textsc{mini};
\NCfBestofnSingleZeroLegalRepProblems{} problems contain one zero-legal single-call replicate,
while other replicates on those problems supply legal inputs.

\paragraph{CF-fresh oracle audit.}
The three-reference consensus oracle behind the CF-fresh buggy pool (\mainpapersection{6}) was
challenged by an independent LLM panel. A gold is the expected output when at least two accepted
human reference solutions run successfully and all successful outputs are identical; an input
whose references produce no such agreement has no gold, is skipped rather than passed, and is
outside the audit by construction. The audited population comprises \NCfOracleGoldUsable{}
agent-arm inputs carrying a usable gold and excludes baseline-arm inputs; a flag is raised when
the panel of sample-passing LLM solutions assigns more votes to some other output than to the
gold. All \NCfOracleContestedTotal{} flags adjudicate to correlated-LLM-error false alarms. Brute-force ground
truth confirms all \NCfOracleContestedTop{} flagged golds on one problem (cf2220F). The remaining
\NCfOracleContestedOther{} flags (cf2231F, whose inputs are too large to brute-force) adjudicate to
the same pattern on structural grounds: the gold is three system-test-passing human solutions
against a sample-passing LLM plurality on a deterministic-answer problem. No demonstrated wrong
gold remains.

\begin{figure}[t]
\centering
\providecommand{\appendixplotwidth}{\columnwidth}
\includegraphics[width=\appendixplotwidth]{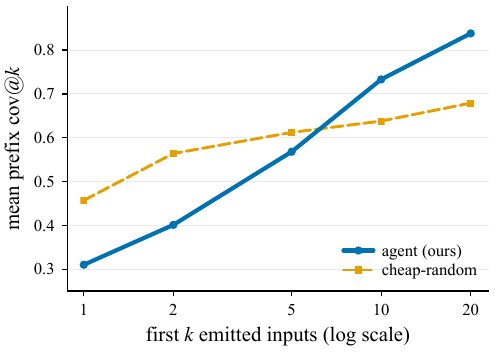}
\caption{Saturation on the TCE \NTceProblems-problem sweep. Markers show absolute mean prefix
cov@$k$---the fraction of each problem's known-bug submissions killed after the first $k$ inputs
in emission order---with problems weighted equally, rather than using a random $k$-subset.
Cheap-random generation matched to the agent's input budget front-loads coverage, while the
agent set continues adding
coverage through the tail. The sweep's known-bug pool is independent
of every compared arm, unlike the CF-fresh pool, which is the union of the arms' own certified
findings. Twenty inputs form the sweep's per-arm design budget, so cov@20 is each arm's
full-budget ceiling.}
\label{fig:tce-saturation}
\end{figure}

\paragraph{TCE saturation sweep.}
The mechanism analysis runs on a \NTceProblems-problem stratified Codeforces sweep spanning five
difficulty bands (\NTceBandLo--\NTceBandHiLo{} rating), built on the TestCase-Eval known-bug
setting \citep{yang-can-2025}. We measure saturation shape against cheap-random generation matched to the agent's input budget:
the normalized coverage at input 1 and the per-input marginal coverage in the saturated tail
(Figure~\ref{fig:tce-saturation} shows the coverage shape; Table~\ref{tab:tce-detail} reports
both statistics). The agent set reaches \NTceAgentCovTwenty{} cov@20, and its mean prefix curve
rises across both measured tail intervals; cheap-random front-loads and then flattens. A no-feedback
configuration, which withholds our model-generated wrong-candidate kill signal, retains almost
the same tail marginal (\NTceMargAgentNofb{} vs \NTceMargAgent{}). The sustained tail marginal is
thus a property of the agent-constructed set, not an artifact of our feedback loop. The ablation rules
out our loop as the source, and claims nothing broader about feedback.

\begin{table}[t]
\centering
\setlength{\tabcolsep}{6pt}
\begin{tabular}{lcc}
\toprule
configuration & norm.\ cov@1 & tail marginal (SE) \\
\midrule
cheap-random       & \NTceSatOneRandom & \NTceMargRandom{} (\NTceMargRandomSe) \\
agent              & \NTceSatOneAgent  & \NTceMargAgent{} (\NTceMargAgentSe) \\
agent, no feedback & ---               & \NTceMargAgentNofb{} (\NTceMargAgentNofbSe) \\
\bottomrule
\end{tabular}
\caption{TCE saturation detail over the \NTceProblems-problem sweep. Normalized cov@1 is
cov@1/cov@20 (higher = more front-loaded), with the denominator taken at the sweep's per-arm
design budget; the tail marginal is mean new coverage per added input over the window from the
sixth input through the smaller of the arm's actual input count and that budget, so arms with
fewer inputs than that budget are not diluted (standard error in parentheses). Both columns average
per-problem values, whereas Figure~\ref{fig:tce-saturation} plots mean coverage curves, so neither
column can be recovered as a ratio of points on those curves. Cheap-random
front-loads and flattens; the
agent set spreads coverage across inputs, and withholding feedback barely moves the marginal.}
\label{tab:tce-detail}
\end{table}

\paragraph{Released tables.}
The per-problem kill matrices for both sweeps (cov@$k$ per problem and per arm, and the TCE
per-problem marginals) are released as machine-checkable tables alongside the buggy pools and
the validator gates. Every cell above is re-derivable.

\paragraph{Artifact availability.}
The supplementary code and data package carries the ledgers, kill matrices, validators, judging
harness, and analysis code behind the numbers reported here, together with the signed
false-discovery adjudication of \S A and the per-finding exhibit bundle.
It ships one entry point that recomputes every number this paper cites from the packaged
artifacts alone and diffs each against the frozen values the paper is typeset from; the
remaining registered measures are retired ones no section cites. A manifest records a SHA-256
digest for every file. What the package re-derives is every reported number from the stored
artifacts. Regenerating a kill matrix one level further back, from the raw submissions, needs
those submissions, and platform terms keep them out of the release; the regeneration script and
the pinned invariant it self-checks against ship anyway, so the procedure is inspectable even
where it is not runnable from the package alone. That path is Linux-only by design: AtCoder
judging requires an unlimited process stack, and the macOS fallback silently caps it, turning
deep-recursion accepted solutions into false runtime errors. The script refuses to run off Linux
rather than emit a plausible wrong count.

\paragraph{Computing environment.}
All authoritative judging runs on one fixed machine: Ubuntu 22.04.3 LTS, Linux 6.2.0-37, x86-64,
an 8-thread Intel Xeon E5-2682 v4 at 2.50\,GHz with 15\,GiB of memory, g++ 11.4.0 (C++17) and
Python 3.10.12. The engines are the released command-line agents and base models cited in
\mainpapersection{3}; each runs against its own provider's API, so wall-clock and token costs
(\S A) are not machine-bound. Hardware affects one verdict class only. Timeouts depend on
processor speed, which is why the reported detection figures use the excl-TLE basis defined in
\mainpapersection{4}: wrong-answer and runtime-error verdicts are deterministic under a fixed stack
limit and reproduce across runs, whereas time-limit verdicts would not.

\section{Qualitative Exhibits}
\label{app:exhibits}

This appendix makes the certified findings concrete: the distribution of what goes wrong, three
transcribed cases, and how the audit corrects itself. Every exhibit is transcribed from a real
released artifact, never synthesized.

\paragraph{What a certified finding is.}
The \NAtcoderAcbBuggy{} verified findings split into \NAtcoderAcbWa{} wrong-answer and
\NAtcoderAcbRe{} runtime-error submissions over \NAtcoderAcbProblems{} problems. A wrong-answer
finding is a submission AtCoder accepted whose output disagrees with the consensus oracle on a
legal input: an input inside the constraints the statement promises, certified by the
problem's validator. A runtime-error finding crashes on such an input. The findings are not
concentrated: per-problem single-linkage clustering over comment-stripped source token 5-shingles
at Jaccard $\geq$ 0.8, with identifiers left unchanged, yields \NAtcoderAcbNearclone{}
near-clone-distinct clusters and \NAtcoderAcbFamilies{} (bug class $\times$ problem) families.

\paragraph{Bug-class profile.}
An LLM labeler assigns each finding a bug class; Table~\ref{tab:bugclass} gives the distribution
over the \NAtcoderBugclassDistinct{} distinct classes. Because every one of
the \NAtcoderAcbBuggy{} entries is a submission the official suite accepted and an agent suite
killed, this profile is by construction the profile of what only the agent audit caught.
The labels are descriptive and single-labeler. An independent blind cross-family re-label
agrees on \NAtcoderBugclassAgreementPct\% of a stratified \NAtcoderBugclassAgreementN-entry
sample at this table's grain (Cohen's $\kappa=\NAtcoderBugclassAgreementKappa$), with
disagreement concentrated on algorithm-level class boundaries. The taxonomy is indicative,
not adjudicated, and per-class counts carry no claims. Verified-bug existence (the
\NAtcoderAcbBuggy{} entries) is untouched: kill verdicts are mechanical, only the class names
are soft.

\begin{table}[t]
\centering
\begin{tabular}{lr}
\toprule
bug class & submissions \\
\midrule
special-case missing   & \NAtcoderBugclassSpecialCaseMissing \\
wrong greedy           & \NAtcoderBugclassWrongGreedy \\
off-by-one             & \NAtcoderBugclassOffByOne \\
integer overflow       & \NAtcoderBugclassIntegerOverflow \\
array out-of-bounds    & \NAtcoderBugclassArrayOutOfBounds \\
boundary               & \NAtcoderBugclassBoundary \\
tie-breaking           & \NAtcoderBugclassTieBreaking \\
wrong algorithm        & \NAtcoderBugclassWrongAlgorithm \\
precision              & \NAtcoderBugclassPrecision \\
other ($+\NAtcoderBugclassOtherClasses$ classes) & \NAtcoderBugclassOther \\
\midrule
total ($\NAtcoderBugclassDistinct$ classes) & \NAtcoderAcbBuggy \\
\bottomrule
\end{tabular}
\caption{Bug-class distribution of the \NAtcoderAcbBuggy{} verified accepted-but-buggy findings
(LLM-labeled; nine most common classes, with \emph{other} aggregating the remaining
\NAtcoderBugclassOtherClasses{} classes).}
\label{tab:bugclass}
\end{table}

\paragraph{Construction-strategy taxonomy.}
Complementing the bug-class profile of what the audit caught, the construction-strategy taxonomy
records the strategies with which the \textsc{codex} arm builds its inputs. The taxonomy is
descriptive and single-labeler; we draw no per-strategy inferences.

\begin{table}[t]
\centering
\setlength{\tabcolsep}{4.5pt}
\providecommand{\fitstrategytable}[1]{\resizebox{\columnwidth}{!}{#1}}
\fitstrategytable{%
\begin{tabular}{lrrr}
\toprule
strategy / feature & present & \%agree & $\kappa$ \\
\midrule
\multicolumn{4}{l}{\emph{construction strategies}} \\
semantic-adversarial construction & \NAtcoderTrajSemanticAdv & \NAtcoderTrajAgreeSemanticAdvPct & \NAtcoderTrajKappaSemanticAdv\textsuperscript{$\dagger$} \\
structural / extreme families & \NAtcoderTrajStructural & \NAtcoderTrajAgreeStructuralPct & \NAtcoderTrajKappaStructural\textsuperscript{$\dagger$} \\
generator sweep & \NAtcoderTrajSweep & \NAtcoderTrajAgreeSweepPct & \NAtcoderTrajKappaSweep \\
small-case enumeration & \NAtcoderTrajEnum & \NAtcoderTrajAgreeEnumPct & \NAtcoderTrajKappaEnum \\
$\geq$2 of the 4 co-occur & \NAtcoderTrajMultiTwo & --- & --- \\
\midrule
\multicolumn{4}{l}{\emph{recurring features}} \\
self-built wrong panel & \NAtcoderTrajSelfWrongPanel & \NAtcoderTrajAgreeSelfWrongPanelPct & \NAtcoderTrajKappaSelfWrongPanel\textsuperscript{$\dagger$} \\
paired differential mining & \NAtcoderTrajPairedMining & \NAtcoderTrajAgreePairedMiningPct & \NAtcoderTrajKappaPairedMining\textsuperscript{$\dagger$} \\
\bottomrule
\end{tabular}}
\caption{Multi-strategy construction in the audited (\textsc{codex}-arm) trajectories: at least two
strategies co-occur in \NAtcoderTrajMultiTwo{} of \NAtcoderProblemsTotal{} trajectories, with one
audited trajectory per problem. \emph{Present} is the pass-1 count over \NAtcoderProblemsTotal{}
problems; strategies are not mutually exclusive. $^\dagger$ marks near-constant rows, where
$\kappa$ degenerates; read \%agree there. Pass-2 does not measure cross-family agreement.
Strategy labels are defined in \S A.}
\label{tab:trajectory}
\end{table}

\paragraph{Three transcribed cases.}
Each case below is one verified finding, transcribed from the released exhibit bundle: a
submission the official suite accepted, the legal killing input that certifies it, and the
oracle's expected output against what the submission produced.
\begin{itemize}
\item \emph{abc167\_e: runtime error.} The submission sizes its factorial tables to
  \texttt{n} (\texttt{vector<llint> mul(n,0)}) but writes \texttt{mul[i+1]} for \texttt{i} up to
  \texttt{k}. On the legal input \texttt{3 2 2}, where \texttt{k} equals \texttt{n-1}, the write
  \texttt{mul[n]} runs off the end and the program crashes; the oracle's expected answer is
  \texttt{8}.
\item \emph{abc161\_f: wrong answer.} Enumerating divisors of \texttt{n} and \texttt{n-1},
  the submission admits a value the statement excludes: on the legal input \texttt{3} it outputs
  \texttt{3} where the oracle requires \texttt{2}.
\item \emph{abc141\_e: wrong answer.} The submission special-cases the first table row
  (\texttt{i=0}) in its own loop, setting \texttt{c[0][j]=1} for a matched pair but omitting the
  \texttt{mx=max(mx,c[i][j])} update that the main loop applies only from the second row onward.
  A length-one repeat found solely in that first row is therefore never counted: on the legal
  input \texttt{2} / \texttt{aa} it outputs \texttt{0} where the oracle requires \texttt{1}.
\end{itemize}
The released exhibit bundle records, for every finding, the LLM labeler's bug class, the killing
input itself, and the observed behavior (the oracle's expected output against what the submission
produced). Problem statements and submission sources stay with the originating platforms; each
record carries the public problem and submission identifiers that locate them.

\paragraph{The audit correcting itself.}
The clearest illustration of the chain's conservatism is a case it \emph{retracted}. On
\texttt{abc164\_e}, deterministic re-judging first flagged \NAtcoderAbcFp{} submissions.
Re-judging each against a 200-submission pool showed the \NAtcoderOracleRefsPerProblem-reference
oracle was itself the outlier on that problem, and all \NAtcoderAbcFp{} were removed from the
ledger. No finding in the released ledger rests on an expectation the larger pool overturns.
The visible count is what survives this correction, not what the first pass produced.

\section{Agent and Single-Call Prompts (Agent-Form Fairness)}
\label{app:prompts}

The agent-form control in the \mainpapername{} (Experiment~2) compares two CF-fresh arms:
\textsc{mini}---not the \textsc{claude} arm in the main results table---and a bare single call
of the same base model, DeepSeek V4 Pro, to separate the agentic workflow from the base model.
The single-call arm was run in four independent reps, whose mean is reported in the \mainpapername{}.
This appendix presents the complete single-call system message and user template, reproduces the
agent-arm task file with one \texttt{[...]} elision, and explicitly excerpts the auditing skill to
which that file points; the full skill file is in the released archive, and
Table~\ref{tab:prompt-equiv} documents the two arms' equivalence so the comparison's fairness is
checkable. The two arms receive the same task objective, the same initial information (the problem
statement and one trusted reference solution), the same $50$-input budget, and the same input-type
checklist. The single-call system prompt was written to mirror the objective and checklist of the
agent skill. Both arms may submit literal inputs or a C++ generator program. For the single-call
arm, the harness compiles the generator and runs it over seeds after the call, treating each run's
standard output as an input, but returns no execution result to the model. The remaining
agent-form differences are construction-time execution feedback and iteration: \textsc{mini} can
check a candidate against the trusted reference, observe the result, and revise before submission,
whereas the single-call arm cannot. The \textsc{mini} arm is nevertheless the no-feedback
ablation, with its weak-coder wrong-solution generator disabled. It receives no kill signal on
which to iterate, so the comparison remains conservative toward the agent.

\begin{table}[h]
\centering
{\small
\setlength{\tabcolsep}{3.5pt}
\begin{tabular}{@{}ll@{}}
\toprule
dimension & equivalent across arms? \\
\midrule
base model                         & yes (DeepSeek V4 Pro) \\
objective + kill definition        & yes \\
initial information (statement{+}reference) & yes \\
legality requirement               & yes \\
$50$-input budget                  & yes \\
input-type checklist               & yes (near-verbatim) \\
generator-program channel          & yes \\
no held-out test pool              & yes \\
kill signal during construction    & yes (none) \\
\midrule
construction-time reference check  & \textbf{no --- agent form} \\
returned execution results         & \textbf{no --- agent form} \\
iteration (revise vs.\ one-shot)   & \textbf{no --- agent form} \\
\bottomrule
\end{tabular}}
\caption{Per-run equivalence of the two arms. Both may emit generator programs and receive no
kill signal; agent form is isolated by access to construction-time checks and their returned
execution results, which permit iterative revision.}
\label{tab:prompt-equiv}
\end{table}

\onecolumn
\subsection{Agent-arm prompt}

Per-problem task identity (\texttt{AGENTS.md}, byte-identical to \texttt{CLAUDE.md}; only the
title, \texttt{docs/statement.txt}, and \texttt{docs/ref.cpp} change across problems):

\begin{promptbox}{Agent-arm task (\texttt{AGENTS.md})}
\begin{lstlisting}[style=promptstyle]
# ADG test-data task -- <pid> (<problem title>)

**Objective.** Build a general discriminating test set -- a diverse
set of legal inputs that catches ANY subtly-wrong solution (the
trusted reference runs OK but the solution diverges -- different
output, TLE, or crash). The goal is the COVERAGE of a diverse,
budget-filling SET -- breadth across distinct failure modes, not a
few clever inputs.

OUTPUT CONTRACT (strict -- this is what gets scored): put your FINAL,
curated test set -- exactly the inputs to be scored, deduplicated, in
the order you want them evaluated -- as numbered files 01.txt ...
50.txt in workspace/final/. Only workspace/final/ is scored.
workspace/hacks/ is scratch.

- How (full protocol + tool contract): follow
  .claude/skills/adg-generator/SKILL.md.
- Problem statement: docs/statement.txt | Oracle: docs/ref.cpp | (no
  validator shipped -- an input is legal if docs/ref.cpp runs OK on
  it; honor the statement's constraints yourself)
- Limits: time 3 s | memory 256 MB
- Input budget -- produce 50 distinct inputs. This is a TARGET to
  FILL, not a ceiling.
- Ablation -- the weak-coder wrong-generator (gen_wrongs.py) is
  DISABLED this run: you have no self-drafted wrongs to probe, so
  there is no kill signal to iterate on. Build from direct reasoning
  about the reference and generator sweeps.

This is an open-ended task -- you choose the strategy and tool order;
nothing is mandatory (see the skill). [...] Do not look for the
held-out evaluation pool: it isn't here, and your only correctness
oracle is docs/ref.cpp.
\end{lstlisting}
\end{promptbox}

The auditing skill this points to (\texttt{.claude/skills/adg-generator/SKILL.md}, objective and
toolbox; excerpted, the full file ships in the released archive):

\begin{promptbox}{Auditing skill \texttt{adg-generator/SKILL.md} (excerpt)}
\begin{lstlisting}[style=promptstyle]
Your objective: produce a high-quality general discriminating test
set for this problem -- a diverse set of legal inputs that FILLS the
input budget, written as numbered files to workspace/final/, that
would catch ANY subtly-wrong solution [...] Optimize the quality of
the test set, not any fixed procedure.

You decide the strategy. This is a toolbox, NOT a pipeline. [...] A
capable test-setter typically:
- Constructs inputs directly [...] edge cases, boundary/extreme
  values, overflow triggers, degenerate/adversarial structures
  (chains, stars, cliques, all-equal, coprime), worst-case inputs.
- Writes a generator program (random or structured) and sweeps
  seeds -- for breadth and for scale.
- Optionally drafts concrete wrong solutions to discover failure
  modes, then finds inputs exposing them.
- Checks candidates, then merges / deduplicates / minimizes into a
  tight non-redundant set.
- Iterates -- and may revisit any of the above multiple times.

Toolbox (every tool is OPTIONAL):
- docs/ref.cpp -- the trusted reference (oracle).
- scripts/eval_hack.py <input> -- check ONE candidate: is it LEGAL,
  does the reference run OK, and which wrongs in workspace/wrongs/
  it kills.
- scripts/write_generator.py workspace/gen.cpp --seeds N -- compile
  and run a C++ generator across N seeds.
- scripts/gen_wrongs.py --n K -- optionally draft weak-coder
  candidate solutions. (DISABLED this run.)
\end{lstlisting}
\end{promptbox}

\subsection{Single-call-arm prompt}

The bare single call (\texttt{cf\_bestofn.py}) is one frozen \texttt{chat.completions} request with
no interactive tools. The harness compiles and sweeps any submitted generator \emph{after} the
call and returns nothing to the model. Unicode punctuation and symbols are transliterated to ASCII
below. System message:

\begin{promptbox}{Single-call system message}
\begin{lstlisting}[style=promptstyle]
You are generating adversarial TEST DATA for a competitive-
programming problem. Your objective: produce a high-quality,
DIVERSE set of LEGAL inputs (satisfying ALL input constraints
stated in the problem) that would catch ANY subtly-wrong solution
-- an input "kills" a wrong solution when the trusted reference
runs OK on it but the wrong solution diverges (different output)
or times out / crashes. You are a frozen frontier model with NO
hidden test pool and NO interactive tools -- reason ONLY from the
statement and the reference solution given, and answer in one
shot. A good set mixes: small/edge cases (n=1, minimal),
boundary/extreme values, overflow triggers, degenerate/adversarial
structures (chains, stars, all-equal, coprime, ...),
worst-case-complexity (maximum-size) inputs, and problem-specific
tricky special cases.
\end{lstlisting}
\end{promptbox}

User message: the problem statement and reference solution, followed by the two-channel protocol
below (\texttt{\{cap\}}${=}50$; \texttt{\{max\_seeds\}}${=}60$; \texttt{\{cxx\}} and
\texttt{\{std\}} are the compiler and standard recorded in \S B):

\begin{promptbox}{Single-call user message (template)}
\begin{lstlisting}[style=promptstyle]
=== PROBLEM STATEMENT ===
{statement}

=== REFERENCE SOLUTION (trusted oracle, human-accepted, C++) ===
{reference}

=== CHANNELS (two ways to produce inputs; use either or both) ===
- **Direct construction** -- emit the raw stdin bytes of an input
  inside a `===INPUT===` block.
- **Generator program** -- emit a C++17 program inside a
  `===GENERATOR===` block. We compile it with `{cxx} -O2
  -std={std}` and run it once per seed `0..N-1`, where N is the
  `===SEEDS===` value; each run's stdout becomes one input.
  **Contract: the program reads ONE integer seed from stdin and
  prints exactly ONE legal input to stdout, and must produce a
  different input for a different seed.** This is the channel for
  SCALE (maximum-size / worst-case inputs) that the literal
  channel cannot reach -- typing a 10^5-element input as text is
  not feasible, writing a generator is.

(!) You will NOT be told whether the program compiled, what it
printed, whether any input is legal, or how many inputs exist so
far. Nothing is checked back to you. Write code that is correct by
construction and inputs that are legal by construction.

=== OUTPUT PROTOCOL (strict) ===
Reply with ONLY the blocks below -- no prose, no explanation, no
markdown fences around them.
===INPUT===
<raw stdin bytes of ONE test input>
===GENERATOR===
<complete C++17 program: reads one integer seed on stdin, prints
ONE legal input on stdout>
===SEEDS===
<how many seeds to sweep, an integer, at most {max_seeds}>
Blocks may appear in any order and any number of times. A
`===GENERATOR===` block MUST be immediately followed by its
`===SEEDS===` block. Emit nothing outside these blocks.

(!!) **THIS IS YOUR ONLY REPLY.** You get exactly ONE response --
there is no second turn, no follow-up, no chance to add more
later. Everything you emit now IS your entire final test set.
Produce all {cap} distinct inputs in this single reply. Do not
pace yourself, do not defer part of the budget to a later turn, do
not end with an intention to continue -- there is no later turn.
\end{lstlisting}
\end{promptbox}

\end{document}